\documentclass[twocolumn,superscriptaddress,showpacs,preprintnumbers,amsmath,amssymb,showkeys,amsfonts,aps]{revtex4}

\usepackage{graphicx}
\usepackage{dcolumn}
\usepackage{bm}

\begin{document}

\preprint{APS/123-QED}

\title{Measurement of the Deuteron Structure Function $\mathbf F_2$
in the Resonance Region and Evaluation of Its Moments\\}

\newcommand*{\INFNGE}{INFN, Sezione di Genova, 16146 Genova, Italy}
\affiliation{\INFNGE}
\newcommand*{\MOSCOW}{Moscow State University, General Nuclear Physics Institute, 119899 Moscow, Russia}
\affiliation{\MOSCOW}
\newcommand*{\ROMATRE}{Universit\`{a} di ROMA III, 00146 Roma, Italy}
\affiliation{\ROMATRE}
\newcommand*{\ASU}{Arizona State University, Tempe, Arizona 85287-1504}
\affiliation{\ASU}
\newcommand*{\UCLA}{University of California at Los Angeles, Los Angeles, California  90095-1547}
\affiliation{\UCLA}
\newcommand*{\CMU}{Carnegie Mellon University, Pittsburgh, Pennsylvania 15213}
\affiliation{\CMU}
\newcommand*{\CUA}{Catholic University of America, Washington, D.C. 20064}
\affiliation{\CUA}
\newcommand*{\SACLAY}{CEA-Saclay, Service de Physique Nucl\'eaire, F91191 Gif-sur-Yvette,Cedex, France}
\affiliation{\SACLAY}
\newcommand*{\CNU}{Christopher Newport University, Newport News, Virginia 23606}
\affiliation{\CNU}
\newcommand*{\UCONN}{University of Connecticut, Storrs, Connecticut 06269}
\affiliation{\UCONN}
\newcommand*{\ECOSSEE}{Edinburgh University, Edinburgh EH9 3JZ, United Kingdom}
\affiliation{\ECOSSEE}
\newcommand*{\EMMY}{Emmy-Noether Foundation, Germany}
\affiliation{\EMMY}
\newcommand*{\FIU}{Florida International University, Miami, Florida 33199}
\affiliation{\FIU}
\newcommand*{\FSU}{Florida State University, Tallahassee, Florida 32306}
\affiliation{\FSU}
\newcommand*{\GWU}{The George Washington University, Washington, DC 20052}
\affiliation{\GWU}
\newcommand*{\ECOSSEG}{University of Glasgow, Glasgow G12 8QQ, United Kingdom}
\affiliation{\ECOSSEG}
\newcommand*{\ISU}{Idaho State University, Pocatello, Idaho 83209}
\affiliation{\ISU}
\newcommand*{\INFNFR}{INFN, Laboratori Nazionali di Frascati, Frascati, Italy}
\affiliation{\INFNFR}
\newcommand*{\ORSAY}{Institut de Physique Nucleaire ORSAY, Orsay, France}
\affiliation{\ORSAY}
\newcommand*{\BONN}{Institute f\"{u}r Strahlen und Kernphysik, Universit\"{a}t Bonn, Germany}
\affiliation{\BONN}
\newcommand*{\ITEP}{Institute of Theoretical and Experimental Physics, Moscow, 117259, Russia}
\affiliation{\ITEP}
\newcommand*{\JMU}{James Madison University, Harrisonburg, Virginia 22807}
\affiliation{\JMU}
\newcommand*{\KYUNGPOOK}{Kyungpook National University, Daegu 702-701, South Korea}
\affiliation{\KYUNGPOOK}
\newcommand*{\MIT}{Massachusetts Institute of Technology, Cambridge, Massachusetts  02139-4307}
\affiliation{\MIT}
\newcommand*{\UMASS}{University of Massachusetts, Amherst, Massachusetts  01003}
\affiliation{\UMASS}
\newcommand*{\UNH}{University of New Hampshire, Durham, New Hampshire 03824-3568}
\affiliation{\UNH}
\newcommand*{\NSU}{Norfolk State University, Norfolk, Virginia 23504}
\affiliation{\NSU}
\newcommand*{\OHIOU}{Ohio University, Athens, Ohio  45701}
\affiliation{\OHIOU}
\newcommand*{\ODU}{Old Dominion University, Norfolk, Virginia 23529}
\affiliation{\ODU}
\newcommand*{\PITT}{University of Pittsburgh, Pittsburgh, Pennsylvania 15260}
\affiliation{\PITT}
\newcommand*{\RPI}{Rensselaer Polytechnic Institute, Troy, New York 12180-3590}
\affiliation{\RPI}
\newcommand*{\RICE}{Rice University, Houston, Texas 77005-1892}
\affiliation{\RICE}
\newcommand*{\URICH}{University of Richmond, Richmond, Virginia 23173}
\affiliation{\URICH}
\newcommand*{\SCAROLINA}{University of South Carolina, Columbia, South Carolina 29208}
\affiliation{\SCAROLINA}
\newcommand*{\JLAB}{Thomas Jefferson National Accelerator Facility, Newport News, Virginia 23606}
\affiliation{\JLAB}
\newcommand*{\UNIONC}{Union College, Schenectady, NY 12308}
\affiliation{\UNIONC}
\newcommand*{\VT}{Virginia Polytechnic Institute and State University, Blacksburg, Virginia   24061-0435}
\affiliation{\VT}
\newcommand*{\VIRGINIA}{University of Virginia, Charlottesville, Virginia 22901}
\affiliation{\VIRGINIA}
\newcommand*{\WM}{College of William and Mary, Williamsburg, Virginia 23187-8795}
\affiliation{\WM}
\newcommand*{\YEREVAN}{Yerevan Physics Institute, 375036 Yerevan, Armenia}
\affiliation{\YEREVAN}
\newcommand*{\deceased}{Deceased}
\newcommand*{\NOWOHIOU}{Ohio University, Athens, Ohio  45701}
\newcommand*{\NOWINDSTRA}{Systems Planning and Analysis, Alexandria, Virginia 22311}
\newcommand*{\NOWUNH}{University of New Hampshire, Durham, New Hampshire 03824-3568}
\newcommand*{\NOWMOSCOW}{Moscow State University, General Nuclear Physics Institute, 119899 Moscow, Russia}
\newcommand*{\NOWSCAROLINA}{University of South Carolina, Columbia, South Carolina 29208}
\newcommand*{\NOWUMASS}{University of Massachusetts, Amherst, Massachusetts  01003}
\newcommand*{\NOWMIT}{Massachusetts Institute of Technology, Cambridge, Massachusetts  02139-4307}
\newcommand*{\NOWURICH}{University of Richmond, Richmond, Virginia 23173}
\newcommand*{\NOWODU}{Old Dominion University, Norfolk, Virginia 23529}
\newcommand*{\NOWGEISSEN}{Physikalisches Institut der Universitaet Giessen, 35392 Giessen, Germany}
\newcommand*{\NOWNONE}{unknown, }

\author {M.~Osipenko} 
\affiliation{\INFNGE}
\affiliation{\MOSCOW}
\author {G.~Ricco} 
\affiliation{\INFNGE}
\author {S.~Simula} 
\affiliation{\ROMATRE}
\author {M.~Battaglieri} 
\affiliation{\INFNGE}
\author {M.~Ripani} 
\affiliation{\INFNGE}
\author {G.~Adams} 
\affiliation{\RPI}
\author {P.~Ambrozewicz} 
\affiliation{\FIU}
\author {M.~Anghinolfi} 
\affiliation{\INFNGE}
\author {B.~Asavapibhop} 
\affiliation{\UMASS}
\author {G.~Asryan} 
\affiliation{\YEREVAN}
% ...
\collaboration{The CLAS Collaboration}
     \noaffiliation
%

\date{\today}

\begin{abstract}
Inclusive electron scattering off the deuteron has been measured
to extract the deuteron structure function $F_2$ 
with the CEBAF Large Acceptance Spectrometer (CLAS) at the Thomas Jefferson National
Accelerator Facility. The measurement covers the entire
resonance region from the quasi-elastic peak up to the invariant mass
of the final-state hadronic system $W\approx 2.7$~GeV with four-momentum
transfers $Q^2$ from $0.4$ to $6$~(GeV/c)$^2$. These data are complementary
to previous measurements of the proton structure function $F_2$ and
cover a similar two-dimensional region of $Q^2$ and Bjorken variable $x$.
Determination of the deuteron $F_2$ over a large $x$ interval
including the quasi-elastic peak
as a function of $Q^2$, together with the other world data,
permit a direct evaluation of the structure
function moments for the first time.
By fitting the $Q^2$ evolution of these moments with an OPE-based twist expansion
we have obtained a separation of the leading twist and higher twist terms.
The observed $Q^2$ behaviour of the higher twist contribution suggests
a partial cancellation of different higher twists entering into the expansion
with opposite signs. This cancellation, found also in the proton
moments, is a manifestation of the ``duality'' phenomenon in the $F_2$
structure function.
\end{abstract}

\pacs{12.38.Cy, 12.38.Lg, 12.38.Qk, 13.60.Hb}

\keywords{moments, deuteron, nucleon structure, higher twists, QCD, OPE}

\maketitle

\section{\label{sec:introduction}Introduction}
Inclusive lepton scattering off the deuteron has provided
a wealth of information about internal nucleon structure
and nuclear phenomena. Since a free neutron target does not exist,
the deuteron is the simplest target for the study of neutron structure
functions. The weak coupling between the two nucleons in the deuteron,
corresponding to a large space-time separation, suggests that
the nucleus can be described as a non-relativistic proton and neutron moving
inside some mean potential.
To this end, the momentum distribution of the deuteron was
established with high precision from the quasi-elastic and $NN$ reactions,
and approaches describing the Fermi motion of the nucleons were developed.
This naive picture, however, was superseded by experiments when the European Muon
Collaboration (EMC) discovered deviations of the measured nuclear structure function $F_2$
from that of a free proton and neutron convoluted with Fermi smearing (a phenomenon
known as the EMC-effect~\cite{EMC-effect}). Different attempts to explain
the EMC-effect have been undertaken, but without reaching a unified
and, therefore, definitive description.

A Quantum Chromodynamics (QCD)-based approach, on the other hand, can handle the nuclear structure
functions in a model-independent manner.
This method is based on the Operator Product Expansion (OPE)
of the structure function moments, whose $Q^2$-evolution is known
in QCD at leading twist and some fixed order in $\alpha_S$.
Note that QCD predictions on the $Q^2$-evolution are target-independent
and nuclear effects are only to modify the normalization of moments,
known in few cases from Lattice QCD simulations.
However, the leading twist picture is valid only in
Deep Inelastic Scattering (DIS) and for not very
large $x$ values. Beyond these kinematic boundaries, e.g. in the region
of large $x$ and moderate $Q^2$, new poorly established physics appears.
This kinematic domain has a particular interest because
multi-parton correlation phenomena manifest themselves as deviations from
perturbative QCD (pQCD) predictions.
For a nuclear target some part of these deviations will be related
to the nucleon interaction with its environment.

In the one photon exchange approximation, the cross section for inclusive electron
scattering off a nucleus is described by the total absorption of the virtual photon
by the nucleus.
The optical theorem relates the total virtual photoabsorption cross section
to the forward Compton scattering amplitude of the virtual photon on the nucleus.
The latter amplitude, in general, can be represented as a product of
two hadronic currents separated by a certain space-time interval $\zeta^2$.
In the Bjorken limit, the interval $\zeta^2 \rightarrow 0$ (while the light-cone
$\zeta^-$ component is fixed) allows one to apply the OPE to the product of
non-local hadronic currents. This leads to a relation where
the moments $M^{CN}_n$ (Cornwall-Norton definition~\cite{Corn-Nort})
of the nucleon (nucleus with mass $A$) structure functions, defined as:
\begin{equation}
M^{CN}_n(Q^2)=\int_0^A dx x^{(n-2)}F_2(x,Q^2),~~~\mbox{$n\ge2$, $n$ even},
\label{eq:CWMoment}
\end{equation}
\noindent are expanded as a series of inverse powers (twists) of the
four-momentum transfer $Q^2$ (for details see Ref.~\cite{Roberts}).
A study of the $Q^2$ dependence of these moments, therefore, would permit one
to isolate different terms of this series, each representing distinct physical
processes in QCD. The first term in the expansion represents the leading twist,
i.e. the limit of asymptotic freedom, while higher terms account
for the interactions among partons inside the nucleon. The contribution
of these multi-parton correlations to the nucleon wave function
increases in the region of large $x$ (corresponding to high moment order $n$)
and low $Q^2$.

For the proton, a careful study of the multi-parton correlation
contribution, which included a global analysis of all the world's data on the
proton structure function $F_2$, has been recently described in
Ref.~\cite{osipenko_f2p}. This analysis could not be done
for the deuteron structure function because of the lack of data in the
region of the quasi-elastic peak, and because of the scarcity of
data in the resonance region. However, a previous
analysis from Ref.~\cite{Ricco2}, based on fits of the structure function
$F_2$, showed a modification of the scaling behaviour of the nucleon structure
function $F_2$ in the nuclear medium.
The Hall~C Collaboration at Jefferson Lab has
recently provided high quality data in this kinematic region~\cite{f2-hc},
but the exclusion of the quasi-elastic peak in this measurement
prevented further studies of these data in terms of QCD.

\begin{figure}
\includegraphics[bb=1cm 6cm 20cm 23cm, scale=0.4]{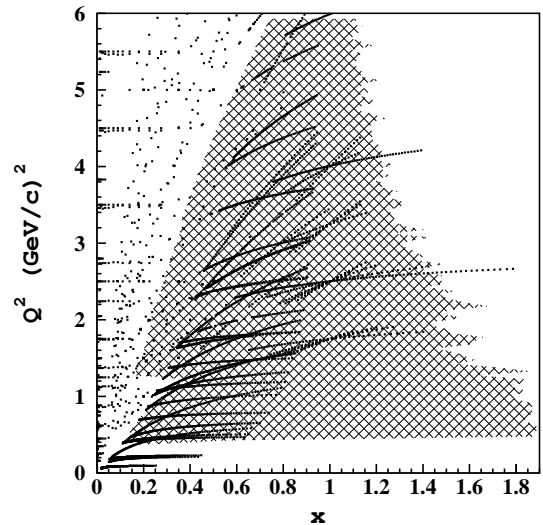}
\caption{\label{fig:xandQ2Domain} Experimental data on the deuteron structure
function $F_2(x,Q^2)$ used for the moment evaluation in the CLAS
kinematic region. The points show world data from
Refs.~\cite{f2-hc,f2-hc_qe,E133,E140,E140x,E49a,E49b,E61,E891,E892,NE11,BCDMS,NMC,NMC_f2d_f2p_ratio,E665,E665_f2d_f2p_ratio}.
The shaded area shows CLAS data.}
\end{figure}

In this paper we report on a measurement of unpolarized inclusive electron
scattering from deuterium taken with the CLAS detector in Hall B at Jefferson Lab.
The data span a wide continuous two-dimensional region
in $x$ and $Q^2$ (see Fig.~\ref{fig:xandQ2Domain}).
The $F_2$ structure function of the deuteron was extracted over the entire
resonance region ($W \le 2.7$~GeV) below $Q^2 = 6$~(GeV/c)$^2$. This
measurement, together with existing world data, allowed for the first time
the evaluation of the first four $F_2$ moments of the deuteron down to
$Q^2 \sim 0.4$ (GeV/c)$^2$.

In section~\ref{sec:Moments} we review the $F_2$ moments in the framework of pQCD.
In section~\ref{sec:DataAnalysis} we discuss improvements in the data analysis
implemented since the first unpolarized inclusive measurement at CLAS,
along with some details of the evaluation of the moments.
For other details of the analysis we refer to Ref.~\cite{osipenko_f2p}.
Finally, in
Section~\ref{sec:Discussion} we discuss the interpretation of the results.

\section{Moments of the Structure Function $F_2$}\label{sec:Moments}
Measured structure functions for a free nucleon target in the DIS
regime are related to parton momentum distributions of the nucleon.
For a nuclear target this relation is not direct since it is necessary
to account for effects of
the Fermi motion, meson exchange currents, off-shellness of the nucleon
and final state interactions (FSI). Nevertheless,
the OPE of the structure function moments of the nuclear structure function
is still applicable in the same way as for the free nucleon.
The $n$-th Cornwall-Norton moment~\cite{Corn-Nort}
of the (asymptotic) structure function $F_2(x,Q^2)$ for a massless target
can be expanded as:
\begin{equation}
M^{CN}_n(Q^2)=\sum_{\tau=2k}^{\infty}E_{n \tau}(\mu_r,\mu_f,Q^2)
O_{n \tau}(\mu_f)\biggl(\frac{\mu^2}{Q^2}\biggr)^{{1 \over 2}(\tau-2)},
\label{eq:i_m1}
\end{equation}
\noindent
where $k=1,2,...,\infty$, $\mu_f$ ($\mu_r$) is the factorization (renormalization)
scale$^1$\footnotetext[1]{We are working in the Soft Gluon Re-summation (SGR)
scheme\cite{SGR}, where $\mu_f^2=\mu_r^2=Q^2$.},
$\mu$ is an arbitrary reference scale,
$O_{n \tau}(\mu_r)$ is the reduced matrix element of the local operators
with definite spin $n$ and twist $\tau$ (dimension minus spin) which is
related to
the non-perturbative structure of the target.  $E_{n \tau}(\mu_r,\mu_f,Q^2)$
is a dimensionless coefficient function describing the small distance behaviour,
which can be perturbatively expressed as a power expansion of the running
coupling constant $\alpha_s(Q^2)$.
Moreover, the leading twist ($\tau=2$) $Q^2$ dependence remains unchanged
with respect to the free nucleon target
and all the nuclear effects appear either in the higher twist terms ($\tau > 2$)
or in the reduced matrix element $O_{n 2}(\mu_r)$, which does not depend on $Q^2$.
The non-zero mass of the target$^2$\footnotetext[2]{In the leading twist approximation
the target is a nucleon inside the deuteron.} leads to additional $M^2/Q^2$
power corrections (kinematic twists) which mix operators of different spin.
These target mass corrections can be accounted for by use of 
Nachtmann~\cite{Nachtmann} moments $M^{N}_n(Q^2)$ instead of the usual
(massless) Cornwall-Norton moments. In the Bjorken limit $M^2/Q^2$
terms become negligible and both definitions coincide.
The Nachtmann moments for the deuteron structure function are defined as
follows:
\begin{eqnarray}\nonumber
M^{N}_n(Q^2) = &&\int_0^2 dx \frac{\xi^{n+1}}{x^3} F_2(x,Q^2)\\ 
&&\Biggl[\frac{3+3(n+1)r+n(n+2)r^2}{(n+2)(n+3)}\Biggr],
\label{eq:i_nm1}
\end{eqnarray}
\noindent
where $r = \sqrt{1+4M^2x^2/Q^2}$, $M$ is the proton mass and $\xi = 2x/(1+r)$.

The evolution of the leading twist term is known 
for the first four moments up to Next-to-Next-to-Leading Order (NNLO).
However, if one wants to extend the
analysis down to $Q^2 \approx M^2$ and to large $x$,
where the rest of the perturbative series becomes significant,
one needs to account for additional logarithmic corrections due to soft gluon
radiation~\cite{SGR,SIM00}. These corrections re-summed in the moment space
to all orders of $\alpha_S$ appear due to an imbalance of the virtual and
real gluon emission at $x\rightarrow 1$.
Since the $Q^2$-evolution of the higher twist terms, related to quark-quark
and quark-gluon correlations, is unknown, their logarithmic QCD
behaviour is parameterized and the corresponding anomalous dimensions
are extracted from the data.

Measurement of the Nachtmann moments $M^{N}_n(Q^2)$ in the intermediate $Q^2$ range
($0.5<Q^2<10$~(GeV/c)$^2$) allows a model-independent separation of the
total higher twist contribution from the leading twist. Comparison of
the higher twist contribution in the deuteron to that in the free proton
provides important insight into the nucleon structure modifications
inside nuclear matter.

\section{Data Analysis}\label{sec:DataAnalysis}
The data were collected at Jefferson Lab in Hall B with the CLAS
using a liquid-deuterium target with thickness $0.81$~g/cm$^2$
during the electron beam running periods in
March-April 2000 and January-March 2002.
The average beam-target luminosity for these periods was
6$\times$10$^{33}$ cm$^{-2}$s$^{-1}$.
To maximize the interval in $Q^2$ and $x$,
data were taken at two different electron beam energies:
$E_0=$~2.474 and 5.770 GeV.
The accumulated statistics at the two energies is large enough ($> 10^9$
triggers) to allow for the extraction of the inclusive cross section with a small
statistical uncertainty ($\le$ 5\%) in small $x$ and $Q^2$ bins
($\Delta x=$0.009, $\Delta Q^2=$0.05 (GeV/c)$^2$).

The CLAS is a magnetic spectrometer~\cite{CLAS_paper} based on a six-coil
torus magnet whose field is primarily oriented along the azimuthal
direction. The sectors, located between the magnet coils, are individually
instrumented to form six independent magnetic spectrometers. The particle
detection system includes drift chambers (DC) for track
reconstruction~\cite{dc},
scintillation counters (SC) for time of flight measurements~\cite{sc},
Cherenkov counters (CC) for electron identification~\cite{cc}, and
electromagnetic calorimeters (EC) to measure neutrals and to improve
electron-pion separation~\cite{ec}.  The EC detectors, which have
a granularity defined by triangular cells in a plane perpendicular
to the incoming particles, are used to study the shape of the
electromagnetic shower and
are longitudinally divided into two parts with the inner part acting as
a pre-shower. Charged particles can be detected and identified for momenta
down to 0.2 (GeV/c) and for polar angles between 8$^\circ$ and 142$^\circ$.
The CLAS superconducting coils limit the acceptance 
for charged hadrons from about 80\%
at $\theta=90^{\circ}$ to about 50\% at forward angles ($\theta=20^{\circ}$).
The total angular acceptance for electrons is about 1.5 sr.
Electron momentum resolution is a function of the scattered electron angle
and varies from 0.5\% for $\theta \leq 30^{\circ}$ up to
1-2\% for $\theta > 30^{\circ}$. The angular resolution
is approximately constant, approaching 1~mrad for polar and 4~mrad
for azimuthal angles: the resolution for the momentum transfer
ranges therefore from 0.2 up to 0.5 \%. The scattered electron missing mass ($W$)
resolution was estimated to be 2.5 MeV for a beam energy less than 3 GeV and
about 7 MeV for larger energies.
To study all possible multi-particle states, the acquisition trigger
was configured to require at least one electron candidate in any of
the sectors, where an electron candidate was defined as
the coincidence of a signal in the EC and Cherenkov modules
for any one of the sectors.

The data analysis procedure has been described in detail in
Refs.~\cite{osipenko_f2p,osipenko_CLAS_Note}.
Therefore, in this article we focus on changes and improvements in the analysis.
The most important improvements, leading to a significant reduction
of the estimated systematic uncertainties relative to those of
Ref.~\cite{osipenko_f2p}, are described in the following sections.

\subsection{Electron Identification}\label{sec:ElPid}
The pion contamination observed in Ref.~\cite{osipenko_f2p}
in the electron candidate sample was found to be due to random coincidences
between a pion track measured in the DC and a noise pulse in a CC photomultiplier tube
(PMT) (typically corresponding to one photo-electron).
These coincidences can be greatly reduced
by means of better matching between the CC hits and the measured tracks.

Each CLAS sector consists of 18 CC segments, each containing two PMTs. Therefore,
the probability of a coincidence is the product of probabilities to have
a noise signal in one of 36 PMTs together with a negative pion track in
a time interval, $\Delta t=$150 ns, which corresponds to the trigger window time.
The average CC PMT noise rate $R^{PMT}$ in CLAS
was measured to be $\approx$42 kHz.
For our typical running conditions, the average rate of negatively charged hadrons
within our geometrical EC fiducial cuts (used to ensure the shower is
fully contained with the detector) that have an appropriate EC signal is of the
order of $R^{h-}\approx$2.3 kHz.
This gives an estimate of the possible contamination:
\begin{equation}
R^{1phe}= R^{PMT} R^{h-} \Delta t\approx 15\mbox{ Hz} ~~,
\end{equation}
\noindent which should be compared to the electron rate
$R^{e-}\approx$250 Hz, using the same cuts.
Therefore, the expected contamination is of the order of 6\% overall. In
contrast,
for small momenta $p < 1$ (GeV/c) and large scattering angles, $\theta>30^\circ$,
$R^{h-}\approx$1.7 kHz and $R^{e-}\approx$100 Hz, resulting in 
a contamination of 12\%.

\begin{figure}[!h]
\includegraphics[bb=1cm 6cm 20cm 23cm, scale=0.4]{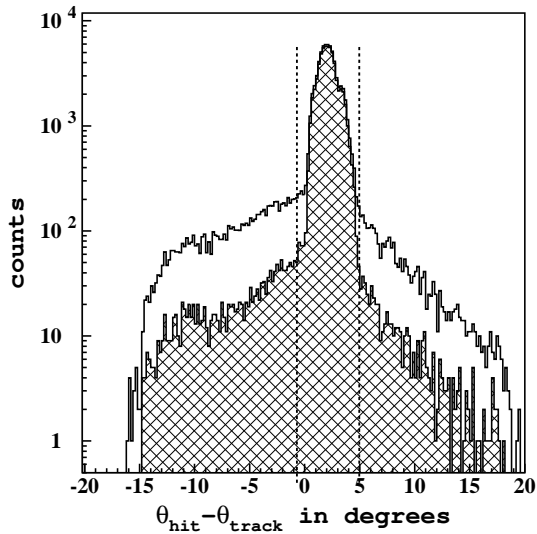}
\caption{\label{fig:thp_cut} The difference in the projective polar angle $\theta$
between the hit position in the CC and the impact point of the electron
candidate track.
The hatched area shows events with reduced pion contamination in the electron
candidate sample. The reduction of the pion contamination is obtained
through an additional cut: the number of photo-electrons in the CC $>2.5$.}
\end{figure}

\begin{figure}[!h]
\includegraphics[bb=1cm 6cm 20cm 23cm, scale=0.4]{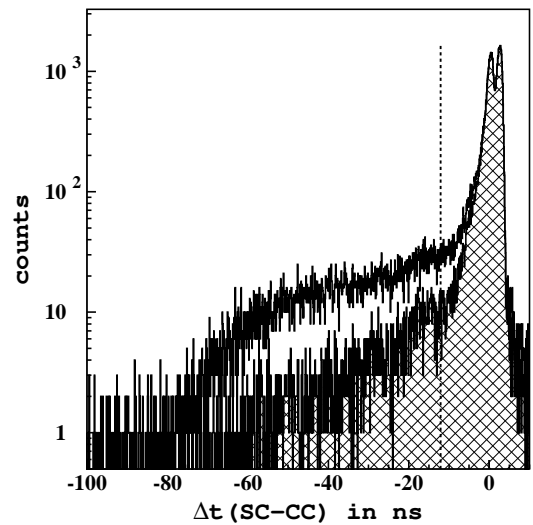}
\caption{\label{fig:dt_cut} The time difference between hits in the SC and
the CC
assigned to the electron candidate track, corrected for the distance traveled
from the SC to the CC.
The meaning of the two histograms is the same as in Fig.~\ref{fig:thp_cut}.}
\end{figure}

\begin{figure}[!h]
\includegraphics[bb=1cm 6cm 20cm 23cm, scale=0.4]{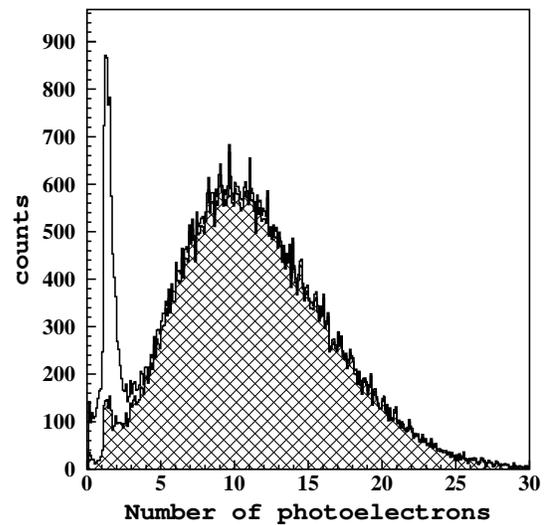}
\caption{\label{fig:after_cut} The spectrum of photo-electrons measured in the
CC. The hatched area represents the CC spectrum after applying the matching
cuts described in the text.}
\end{figure}

In order to reduce the contamination of the coincidences between a
hadron track and a CC noise signal,
we applied geometrical and time matching requirements between the CC signal and the
measured track in the following way:
\begin{enumerate}
\item we defined a CC projective plane, an imaginary plane
behind the CC
detector where Cherenkov radiation would arrive if it were to propagate
the same distance
from the emission point to the PMT without any reflections in the mirror
system;
\item for each CC segment we found the polar angle from the CLAS center to
the image
of the CC segment center and to the images of the CC edges;
\item the impact point and the direction of the track in the SC plane, as
measured in the DC, were
used to obtain the measured polar angle $\theta$ in the projective plane
for each electron candidate event.
We fitted the $\theta$ distributions separately for each CC segment in order
to extract their measured width $\sigma_p$;
\item for each CC segment we applied a cut:
\begin{equation}
|\theta_{track}-\theta_{hit}| < 3 \sigma_p
\end{equation}
which was intended to remove those electron event candidates for which the
track impact point in the CC was far away from the segment where the hit was
detected.
In Fig.~\ref{fig:thp_cut} an example of the $\theta$ distribution
is shown for one segment and the cut applied is indicated by dashed lines.
To clearly identify the contribution of the coincidences and check
the efficiency of the cut we separated events outside the single
photo-electron peak (which contains most of the pion contamination) by
applying a cut, $N_{phe}>2.5$. These electron candidate events with reduced
pion contamination are shown in Fig.~\ref{fig:thp_cut} by the hatched
histogram. The difference between the empty and the hatched histogram in
the figure is therefore mostly due to the pion contamination;
\item to perform time matching between the CC and SC hits of the electron
candidate event we studied the distribution of the time-offset of the CC
signal with respect to the SC. For each CC segment
we measured $\Delta t^{SC-CC}$, defined as follows:
\begin{equation}
\Delta t^{SC-CC} = t^{SC}-t^{CC}-\frac{l^{SC}-l^{CC}}{c\beta} ~~,
\end{equation}
\noindent where $t^{SC}$ and $t^{CC}$ are hit times,
$l^{SC}$ and $l^{CC}$ are the path lengths from the CLAS center to the hit
points in the SC and CC, and $\beta$ is the track velocity measured in the SC.
The distribution of SC-CC time-offsets is shown in Fig.~\ref{fig:dt_cut}.
The electron-rich events outside the single photo-electron peak
are emphasized by the hatched histogram, and the cut applied is shown by
the dashed line.
The presence of the double peak near $\Delta t^{SC-CC} \approx 0$
is expected due to
a small time offset between the two PMTs within the CC segment;
\item The measured photo-electron distribution for the electron candidate
events was compared to the one obtained after all the cuts described above
had been
applied. This comparison is shown by different histograms in
Fig.~\ref{fig:after_cut}.
\end{enumerate}
\noindent
After applying the matching procedure, the pion contamination
was reduced from 30\% to 5\% in the worst case.
The remaining contamination is due to events where the
pion impact point in the CC is very close to the PMT with
the noise signal. This contribution was removed
by the same procedure as in Ref.~\cite{osipenko_f2p}.
In order to estimate the systematic uncertainties of this correction
we compared the method to another approach, namely, requiring more than
2.5 photo-electrons in the CC and estimating the number of missing
electrons by an extrapolation, using the
empirical function from Ref.~\cite{osipenko_cc_matching} adjusted for each
particular run period.
The difference between the two methods provides an estimate
of the systematic uncertainties in the pion rejection corrections.
The total relative systematic uncertainty of this correction 
is kinematic dependent;
at larger $Q^2$ the contribution of pions and the corresponding
systematic uncertainty are larger.

To improve electron identification at large momenta, the energy
deposited in the EC was used. Starting from momenta of 2.7~(GeV/c),
pions begin to emit Cherenkov light in the CC. In this region, however,
the EC becomes very efficient in the separation of electrons from
pions (see Ref.~\cite{ec}), and moreover, the ratio of pion to electron
yields drops down very quickly. Electrons passing through the EC
release 30\% of their energy on average, while pion losses are constant.
This EC property was exploited at large particle momenta for
electron-pion separation by selecting particles with the energy
fraction released in the EC above 20\%. More details on this procedure
can be found in Ref.~\cite{osipenko_f2p}. Furthermore, pions just above the Cherenkov
threshold emit a small amount of light with respect to electrons of
the same momenta, and therefore, were also removed by the cut on the
number of photo-electrons.

\subsection{$e^+e^-$ pair production}
\begin{figure}[!h]
\includegraphics[bb=1cm 6cm 20cm 23cm, scale=0.4]{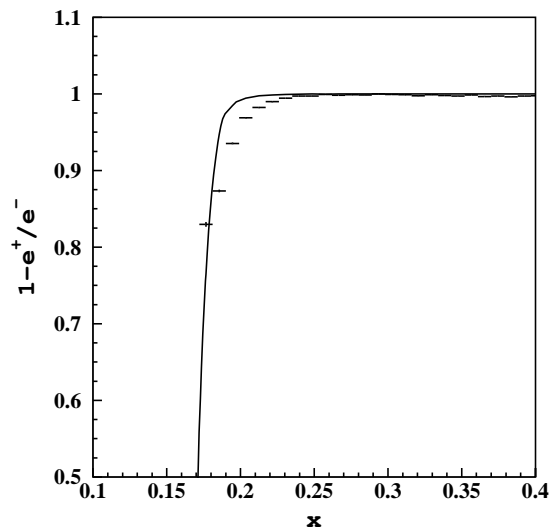}
\caption{\label{fig:epem} The contribution of $e^+e^-$ pair
production events in the inclusive cross section at $Q^2=1.775$ (GeV/c)$^2$.
The points show the measured quantity $1-e^+/e^-$, which represents the
number of electrons inelastically scattered off deuterium to the total
number of measured electrons.
The curve represents the calculations from Ref.~\cite{bosted_epem_eg1}.}
\end{figure}

The most important source of $e^+e^-$ pairs in the CLAS is $\pi^0$ production followed
either by Dalitz decay to $\gamma e^+e^-$ or by $\gamma \gamma$ decay
with one of the photons converting to $e^+e^-$.
For the data set at higher beam energy some measurements were taken
with an out-bending torus field$^3$\footnotetext[3]{Normal setting of the CLAS torus
magnet bends electrons in the forward direction along the beam (i.e. in-bending).
The inverse magnetic field configuration is called out-bending.}.
This provides the possibility to extract
the contribution of $e^+e^-$ pair production directly from the data.
To do so, we demanded the first particle in each event to be
a positron (positron trigger), i.e. to have positive charge
and hits in both the CC and the EC.
We applied exactly the same cuts and corrections described
above and in Ref.~\cite{osipenko_f2p} to the positron
trigger data. Following the
procedure described in Ref.~\cite{bosted_epem_eg1} an additional severe cut
on the number of photo-electrons measured in the CC was applied
($N_{phe}>4$) to both the electron and the positron trigger rates.
The ratio $e^+/e^-$ obtained in this way is shown in Fig.~\ref{fig:epem}
in comparison with calculations described below.
For the higher beam energy data set we therefore subtracted the measured
$e^+e^-$ background accounting for the statistical uncertainties only.

Since positron data are not available for the lower energy data set, the pair
production background processes were estimated
according to a model developed by P.~Bosted and described in
Ref.~\cite{bosted_epem_eg1}. Bosted developed a computer code
based on the Wiser fit of inclusive pion production. This
model was carefully checked against measurements of
the inclusive $e^+/e^-$ ratio in different CLAS runs
on polarized proton and deuteron targets~\cite{bosted_epem_eg1},
and it appeared to be in good agreement (within 30\% relative uncertainty)
with the measured ratio.
The value of the correction is assumed
to be equal to the ratio of the inclusive $e^+$ production cross
section over the fit of the deuteron inclusive cross section $\sigma^M_{rad}$,
including radiative processes
(the tail from the elastic and quasi-elastic peaks, bremsstrahlung and
the Schwinger correction). This correction factor is given by:
\begin{equation}
F_{e^+e^-}(E,x,Q^2)=\frac{
\sigma^M_{rad}(E,x,Q^2)}{
\sigma^M_{rad}(E,x,Q^2)+
\sigma_{e^+}(E,x,Q^2)}   ~~,
\label{eq:d_pp1}
\end{equation}
\noindent
where $\sigma_{e^+}$ is the inclusive $e^+$ production cross
section and $\sigma^M_{rad}$ is the fit folded with radiative processes.
The index ``$M$'' here refers to the model cross section used in the
event generator.

To estimate systematic uncertainties in the calculations, we compared
the calculated $e^+e^-$-pair production contribution to the measured one
for the higher beam energy data. The difference,
shown in Fig.~\ref{fig:epem},
has been parameterized as a function of $y=\nu/E$ and is given by:
\begin{equation}
\delta^{e^+e^-}=0.16 \exp{\Bigl\{-\frac{1}{2}\Bigl(\frac{y-1}{0.1}\Bigr)^2\Bigr\}} ~~,
\end{equation}
\noindent
where $E$ is the beam energy and $\nu$ is the energy of the virtual photon
in the Lab frame.
The systematic uncertainty estimated for the lower beam energy does not exceed 3.5\%.

\subsection{Momentum corrections}
A new correction procedure has been developed in Ref.~\cite{e6a_mom_cor}.
This procedure is based on studying completely exclusive reactions
with overdetermined kinematics in order to correct simultaneously
both momenta and polar angles $\theta$ for all particles (azimuthal angle $\phi$ was
assumed to be correct). The momentum change $\Delta P$ were parameterized as:
\begin{eqnarray}
&& \frac{\Delta P}{P}=
\Biggl\{(E+F\phi)\frac{\cos{\theta}}{\cos{\phi}}+(G+H\phi)\sin{\theta}\Biggr\} \\ \nonumber
&& \frac{P}{q B_{Torus}}+
 (J\cos{\theta}+K\sin{\theta}+L\sin{2\theta})+ \\ \nonumber
&& (M\cos{\theta}+N\sin{\theta}+O\sin{2\theta})\phi ~~~,
\label{eq:d_mcz1}
\end{eqnarray}
\noindent where $q$ is the particle charge (in electron charge units) and
\begin{eqnarray}
&& B_{Torus}=0.76 \frac{I_{Torus}\sin^2{4\theta}}{3375 \theta}~~~~~~\theta < \pi/8  \\ \nonumber
&& B_{Torus}=0.76 \frac{I_{Torus}}{3375 \theta}~~~~~~\theta \geq \pi/8 ~~~.
\end{eqnarray}
\noindent $I_{Torus}$ is the Torus magnet current in A
and the polar angle change $\Delta \theta$ is given by
\begin{equation}
\Delta \theta=(A+B\phi)\frac{\cos{\theta}}{\cos{\phi}}+(C+D\phi)\sin{\theta}
\label{eq:d_mcz2}
\end{equation}
\noindent 14 free parameters ($A-O$) were determined separately 
for each CLAS sector from fitting the sum of the squared differences
of the initial and final four-momentum components in the reactions:
\begin{eqnarray}
&& e + p \rightarrow e^\prime + p^\prime \\ \nonumber
&& e + p \rightarrow e^\prime + p^\prime + \pi^+ \pi^- \\ \nonumber
&& e + d \rightarrow e^\prime + p^\prime + p^{\prime\prime} + \pi^- \nonumber ~~~.
\end{eqnarray}
\noindent 
During the fitting, the hadron momenta were corrected for the energy losses
inside the target, and a better beam energy determination was applied~\cite{beam_en}.

Systematic errors of the momentum correction procedure were studied by means
of a comparison between data and simulations. Simulated spectra were shifted
assuming a constant $\Delta P$ shift that equalized the positions
of the quasi-elastic peaks. The maximum deviation in both sets is
of the order of 5 MeV, which is smaller than the CLAS resolution. 
Also the position of the neutron peak in the $e^\prime\pi^+$ missing mass spectrum
was checked and found to be in agreement with the nominal
neutron mass value within 3 MeV.
The resulting difference in terms of the $F_2$ structure function
between the direct simulation spectra and the shifted ones was
assumed to be an estimate of the corresponding systematic error.

\subsection{Simulations}
Determination of the acceptance and efficiency 
corrections was based entirely on the Monte Carlo (MC) simulations developed for CLAS.
Moreover, the systematic uncertainties of these corrections
were estimated from a comparison of MC simulations with
experimental data using a realistic model in the event generator.
In short, the procedure is the following:
we generated events with the event generator describing
elastic, quasi-elastic and inelastic $eD$-scattering processes including
radiative corrections; these events then were processed with
GEANT based CLAS software simulating
the detector response; after that the standard CLAS event reconstruction procedure
was applied to obtained detector signals; finally the ratio
of reconstructed events to the number of generated
ones gave a combined efficiency/acceptance correction in each kinematic bin.

\begin{figure}
\includegraphics[bb=1cm 6cm 20cm 23cm, scale=0.4]{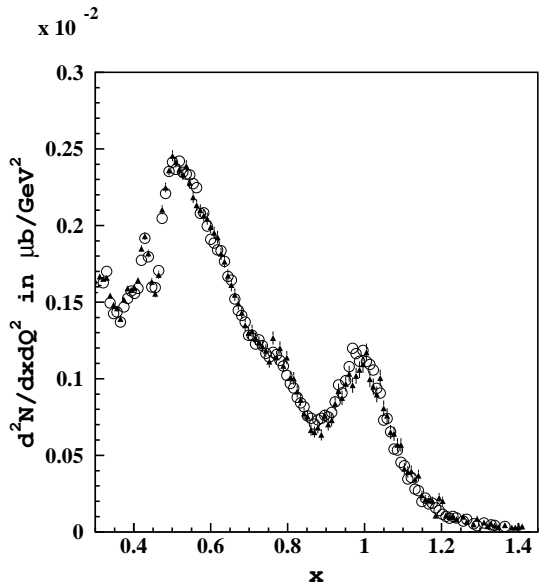}
\caption{\label{fig:raw_cs} The normalized event yield obtained from the
data (filled triangles)
and simulations (open circles) at $E=5.77$ GeV, $Q^2=2.425$ (GeV/c)$^2$.
The yields were obtained within fiducial and EC cuts. An $e^+e^-$
correction was applied to the data. No acceptance or efficiency correction
was applied to either spectrum.}
\end{figure}

\begin{figure}
\includegraphics[bb=1cm 6cm 20cm 23cm, scale=0.4]{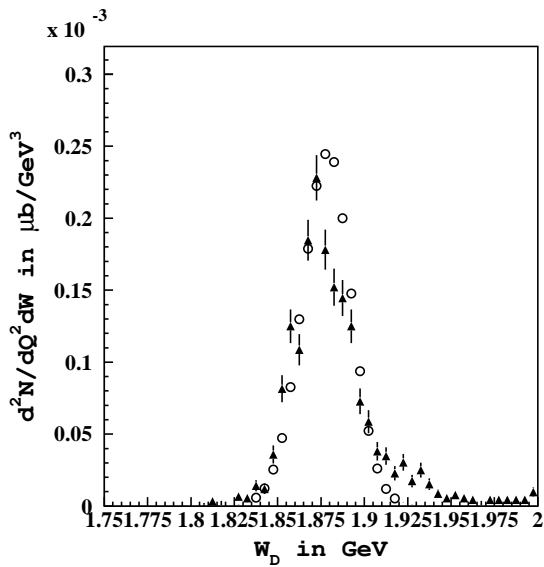}
\caption{\label{fig:SimEl} Normalized yield of elastic electron scattering
off the deuteron at $Q^2=0.47-0.63$ (GeV/c)$^2$: the data are shown by the
triangles and the simulations are shown by the open circles.
No acceptance or efficiency corrections were applied.}
\end{figure}

The simulations of the detector response were performed in the same
way as described in Ref.~\cite{osipenko_f2p}.
The following improvements and changes for electron-deuteron
scattering were implemented:
\begin{enumerate}
\item 
Electron scattering events
were generated by a random event generator with the
probability distributed according to $\sigma^D_{rad}$, described in
Appendix.~\ref{app:f2_model}.
The values for the elastic and inelastic cross sections
for electron-deuteron scattering were taken from existing fits of
world data, in references~\cite{Stuart_ed} and~\cite{f2-hc}, respectively.
The contribution from internal radiative processes was added according to
calculations~\cite{Mo}.
\item The event rate obtained in the simulations was then compared to the data,
preserving the original normalizations (accumulated Faraday Cup (FC) charge
for the data and the number of generated events over the integrated cross
section of the event generator for simulations). These normalized yields
do not include acceptance, efficiency and radiative corrections.
The simulated events passed all cuts: the fiducial
cuts, calorimeter cut, event status cut (see Ref.~\cite{osipenko_f2p} for
details) and CC matching cut. However, $e^+e^-$ pair production
and empty target backgrounds were subtracted from the data.
The normalized yield obtained
with the same set of cuts from the data and simulations were
compared and found to be in good agreement within $\approx 10-15$\%
(see Fig.~\ref{fig:raw_cs}),
which is at the level of reliability for our cross section models.
As one can see below in this section, a $\approx 10-15$\% variation of the
cross section model in the event generator yields only $\approx 1$\%
uncertainty in the final cross section.
\end{enumerate}
\noindent In order to check the absolute normalization of the inclusive $e^-D$ events,
we used elastic $e^-D$ scattering data.
In the range of $Q^2$ considered in this analysis (0.47-0.63 (GeV/c)$^2$) , 
the $e^-D$ elastic cross sections are known within 5\% (see Ref.\cite{GVO_review}).
Recent measurements performed in Jefferson Lab~\cite{halla_ed,hallc_ed}
provide the most precise data in this kinematic region.
We performed the simulations of this reaction
using the parameterization from Ref.~\cite{Stuart_ed} and
taking into account radiative corrections. The normalized event yield
obtained from simulations ($\frac{d\sigma}{d\Omega}_{sim}$)
was compared to the measured one ($\frac{d\sigma}{d\Omega}_{exp}$)
at the lowest $Q^2$ values.
The yields, shown in Fig.~\ref{fig:SimEl}, integrated over the peak,
are in good agreement within
statistical and systematic uncertainties.
The distortion of the measured peak is due to
the $W_D=\sqrt{M_D^2+2M_D\nu-Q^2}$-dependence of the radiative
corrections, which were not taken into account in the simulations.
This $W_D$-dependence is generated by the radiative tail from
the $e^-D$ elastic peak which in our approach to radiative corrections
begins at higher $W_D$ due to the cut on soft photons.
The efficiency obtained from simulations and checked against the elastic
scattering data appear to be approximately constant
(about 90-95\%) inside the fiducial region of the detector defined
by the fiducial cuts.

The elastic scattering normalized yield was evaluated as
the number of $e^-D$ coincidences measured in CLAS in the
$W_D$-interval from
$1.75$ to $2$ GeV multiplied by the corresponding
luminosity:
\begin{equation}
\frac{d\sigma}{d\Omega}_{exp,sim}(E,\theta)=
\frac{L_{sim}}{\rho \frac{N_A}{M_A} L Q_{tot}}
\int_{1.75}^{2}dW_D
N_{exp,sim}(W_D,\theta) ~~,
\label{eq:d_e_ee1}
\end{equation}
\noindent where $N_{exp,sim}$ represent the corresponding numbers of events
(for the measured cross section the empty target events
were subtracted).
In order to clean up the elastic data sample, we
applied additional cuts on the deuteron identification
$|M_D^{2 \mbox{ }(exp)}-M_D^2| < 0.5$ GeV$^2$ and the kinematic correlations
between the electron and deuteron:
\begin{eqnarray}
&& ||\phi_e-\phi_D|-180^\circ| < 3^\circ \mbox{   and } \\ \nonumber
&& \Biggl|\cos{\theta_D}-\frac{(E_0-P_e \cos{\theta_e})}{P_D}\Biggr| < 0.1 ~~,
\end{eqnarray}
\noindent where $M_D^{\mbox{ }(exp)}$ is the mass of the deuteron measured
in the SC; $M_D$ is the nominal deuteron mass; $\phi_e$ ($\theta_e$) and
$\phi_D$ ($\theta_D$) represent the
azimuthal (polar) angles (in degrees) of the scattered electron and deuteron,
respectively;
$P_e$ and $P_D$ are the momenta of the particles. As one can see from
Fig.~\ref{fig:SimEl}
the elastic peak is very well separated from the inelastic background.
The integrated peak yields agree to within the 3\% statistical uncertainty.

There are two systematic uncertainties in the simulation.
The first one is due to the model dependence of the
reaction cross section used for generating the events.
We applied a different cross section model for the inelastic electron-deuteron
scattering, taken from Ref.~\cite{Simula_QE}, and the differences obtained for
the efficiency were taken as an estimate of the systematic
uncertainty. These systematic uncertainties were averaged over both
kinematic variables to give a uniform systematic uncertainty to both data
sets, which was estimated to be 1.7\%. The second systematic uncertainty is due to
the inability of the GEANT3-based CLAS
simulation package GSIM~\cite{GSIM} to perfectly reproduce the CLAS response
to electron tracks at different
angles and momenta. To estimate this effect, we treated the six CLAS sectors
as independent spectrometers. The normalized event yield measured
in each sector was compared separately to the simulations,
as shown in Fig.~\ref{fig:raw_cs}.
The observed differences were compared sector-by-sector to remove
uncertainties due to the event generator model.
From this comparison we obtained a systematic
uncertainty varying from 3 to 6\% depending mostly on the scattered electron
polar angle. The two uncertainties were summed in quadrature.

\subsection{Structure Function $F_2(x,Q^2)$}\label{sec:d_sf}
The measured electron yields $N_{exp}$, normalized to the integrated luminosity 
in conjunction with Monte Carlo simulations, were used to extract
the structure function $F_2$ in each kinematic bin.
The Monte Carlo events were used to simultaneously obtain efficiency,
acceptance, bin centering and radiative corrections. $F_2$ was determined using:
\begin{eqnarray}
&&  F_2(x,Q^2) =
\frac{1}{\rho \frac{N_A}{M_A} L Q_{tot}} 
\frac{J}{\sigma_{Mott}} \frac{\nu}{1+\frac{1-\epsilon}{\epsilon}\frac{1}{1+R}} \nonumber\\
&&\frac{N_{exp}(x,Q^2)}
{\Psi(x,Q^2)}
F_{phe}(x,Q^2)
F_{e^+e^-}(x,Q^2)~,
\label{eq:d_ii2}
\end{eqnarray}
\noindent
where $\rho$ is the density of liquid $D_2$ in the target, $N_A$ is
the Avogadro constant, $M_A$ is the target molar mass, $L$ is the
target length, $Q_{tot}$ is the total charge in the Faraday Cup (FC)
and $\Psi(x,Q^2)$ is the efficiency including the radiative
and bin-centering correction factors:
\begin{equation}
\Psi(x,Q^2)=\Psi_{eff}(x,Q^2)\Psi_{rad}(x,Q^2)
\Psi_{bin}(x,Q^2)~,
\label{eq:d_ii3}
\end{equation}
\noindent
with:
\begin{equation}
\Psi_{rad}=\frac{\sigma^M_{rad}}{\sigma^M}~~~
\mbox{and}~~~\Psi_{bin}
= \frac{\int_{\Delta \tau} d\sigma^M}{\sigma^M}~,
\label{eq:d_ii4}
\end{equation}
\noindent
$\Psi_{eff}$ is the ratio between the number of reconstructed and
generated events in the bin.
The integral in Eq.~\ref{eq:d_ii4} was taken over the current bin area $\Delta \tau$.
Here $\epsilon$ is the virtual photon polarization parameter:
\begin{equation}
\epsilon \equiv \Biggl(
1+2\frac{\nu^2+Q^2}{Q^2} \tan^2{{\theta \over 2}}
\Biggr )^{-1}~.
\label{eq:d_sf2}
\end{equation}
\noindent 
The Mott cross section $\sigma_{Mott}$ and the Jacobian $J$
of the transformation between $d\Omega dE^\prime$ to $dx dQ^2$
are defined by:
\begin{equation}
\sigma_{Mott}=\frac{\alpha^2 \cos^2{\frac{\theta}{2}}}{4E^2\sin^4{\frac{\theta}{2}}}
~~~\mbox{and}~~~
J=\frac{x E E^{\prime}}{\pi \nu}~.
\end{equation}
\noindent 
The structure function $F_2(x,Q^2)$ was extracted using
the fit of the function $R(x,Q^2)\equiv \sigma_L / \sigma_T$
described in Appendix~\ref{app:rlt_model}.
However, the structure function
$F_2$ in the relevant kinematic range is very insensitive
to the value of $R$. For example in typical kinematics
$\epsilon = 0.75$ and assuming SLAC DIS $R$ value of 0.18 the relative
uncertainties of $F_2$ and $R$ are related: $dF_2/F_2=0.03 dR/R$.
Therefore, in this kinematics 20\% error on $R$ will generate
only 0.6\% error on $F_2$.
The overlapped data from two different beam
energies were combined using weighted average technique.
Moreover, we checked that the parameterization used for $R$
is consistent with the difference between two cross sections
within statistical and systematic errors.

\begin{figure*}
\includegraphics[bb=1cm 5cm 19cm 23cm, scale=0.4]{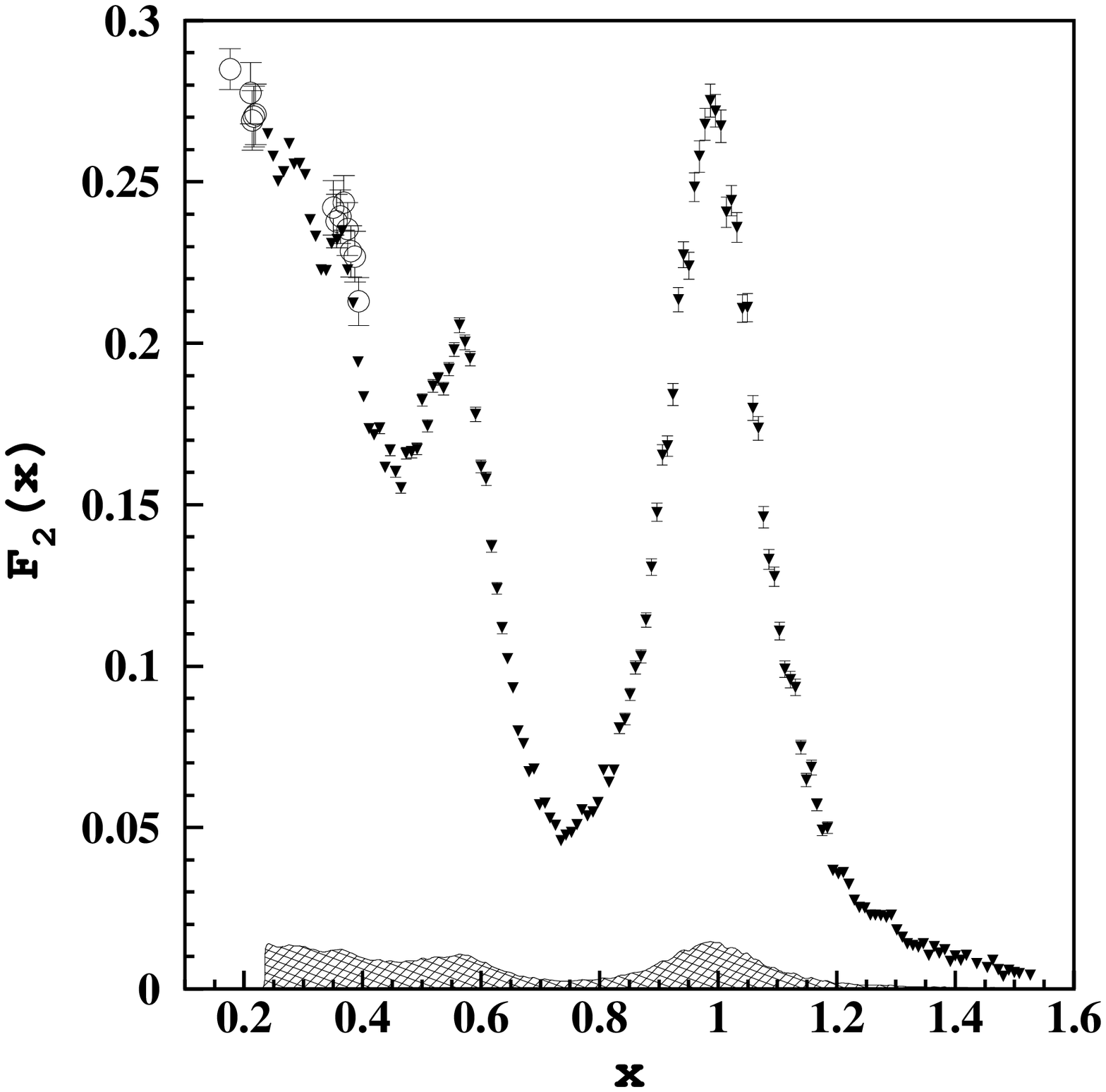}~%
\includegraphics[bb=1cm 5cm 19cm 23cm, scale=0.4]{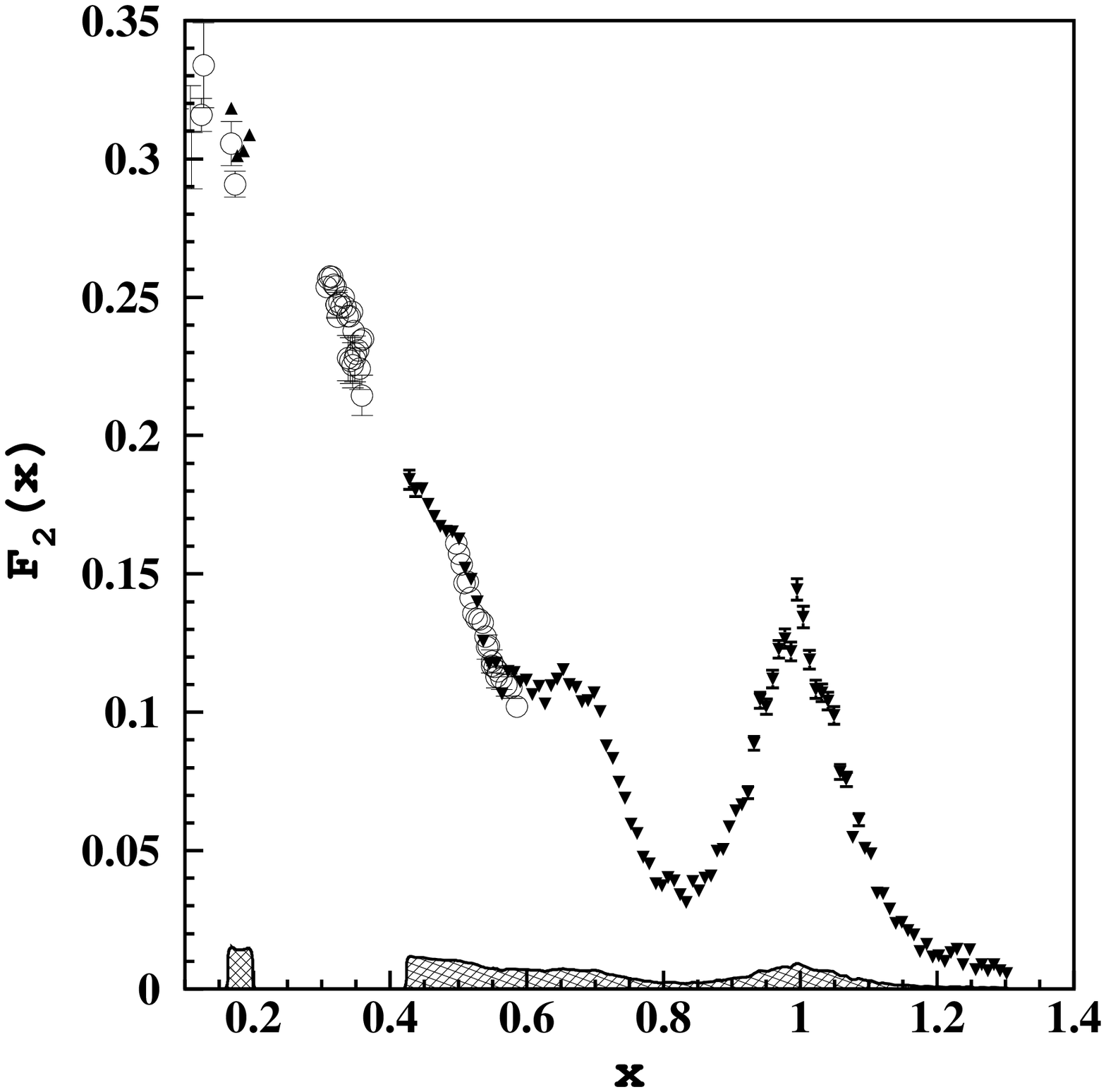}
\includegraphics[bb=1cm 5cm 19cm 23cm, scale=0.4]{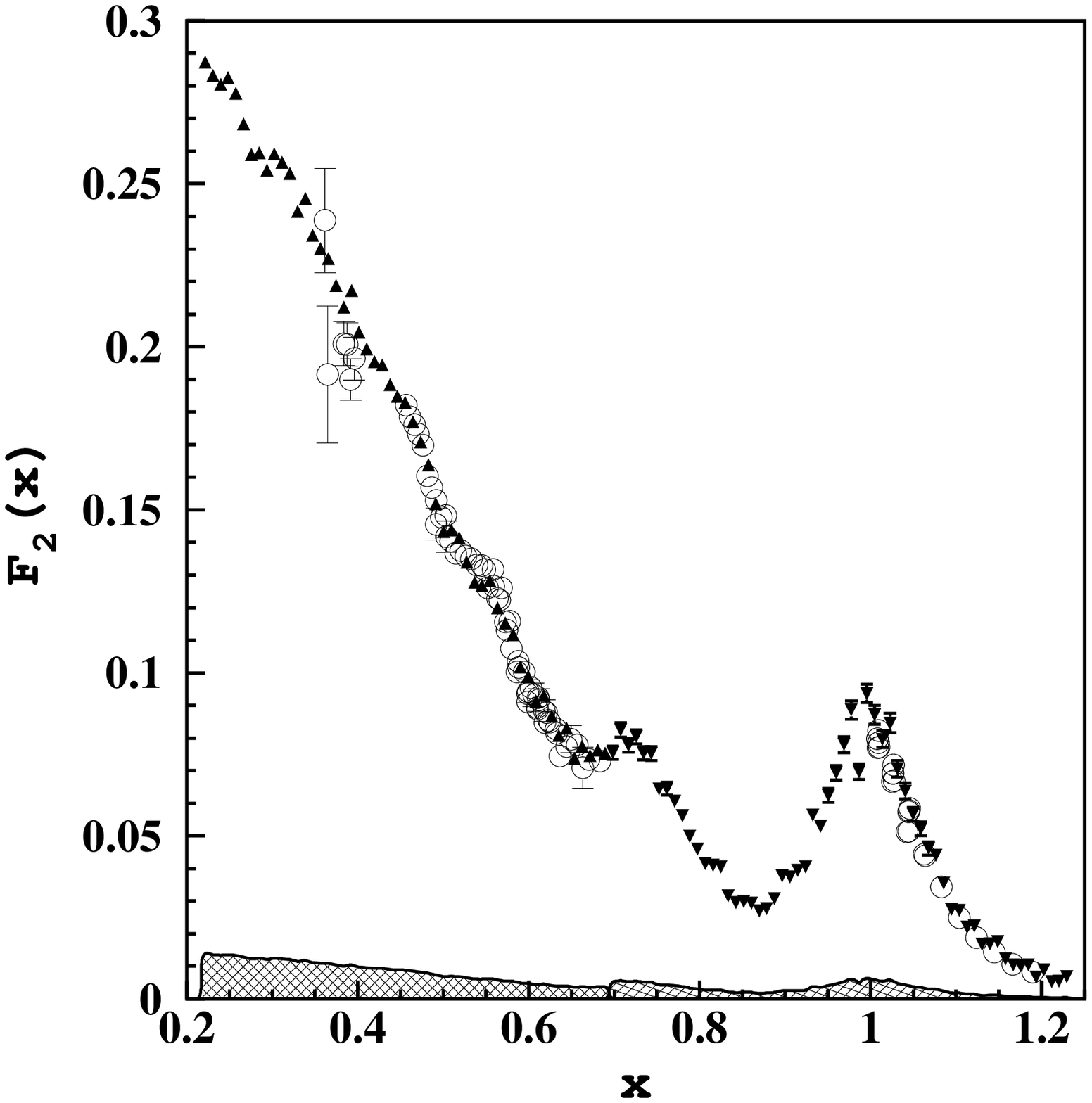}~%
\includegraphics[bb=1cm 5cm 19cm 23cm, scale=0.4]{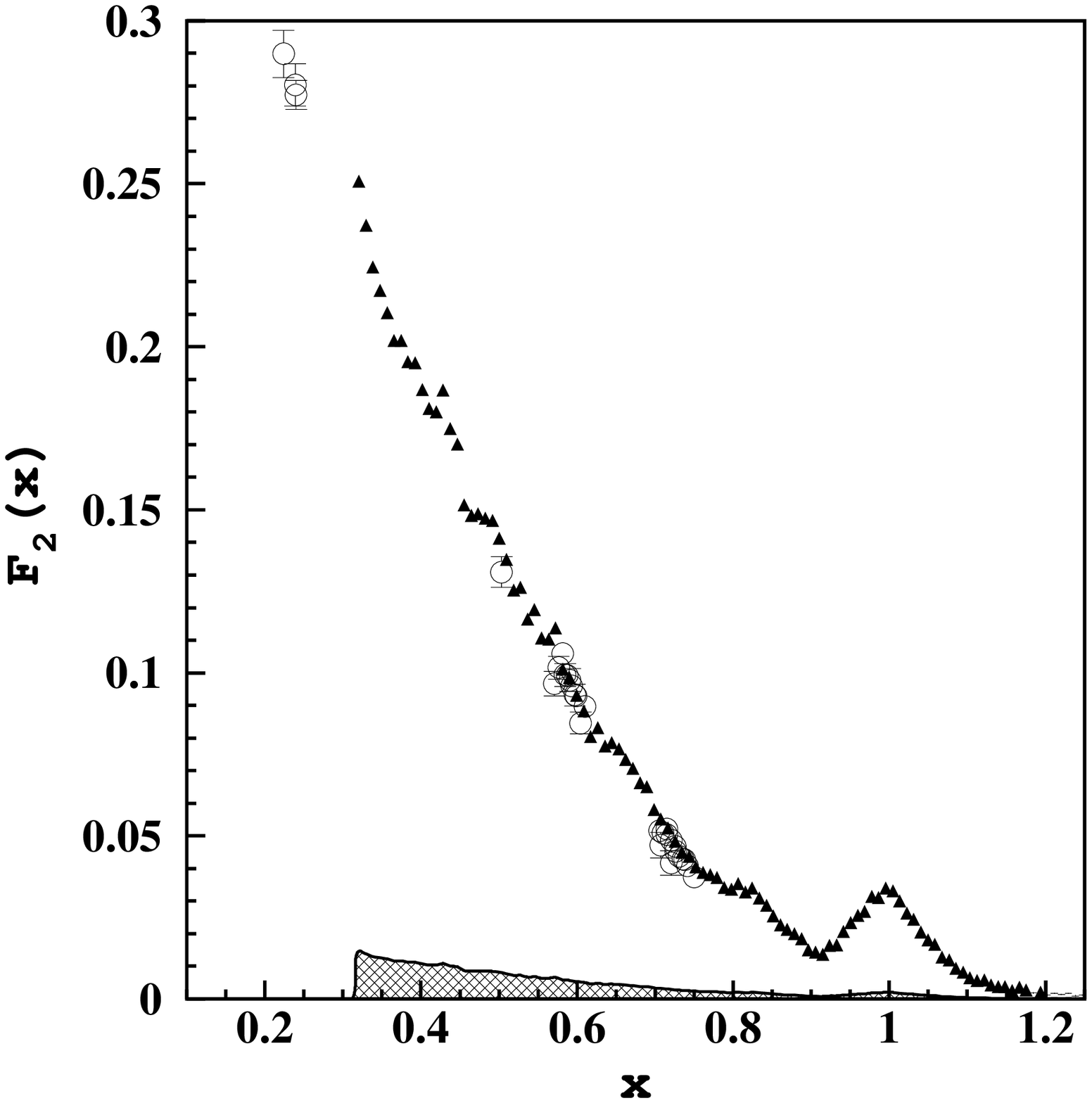}
\includegraphics[bb=1cm 5cm 19cm 23cm, scale=0.4]{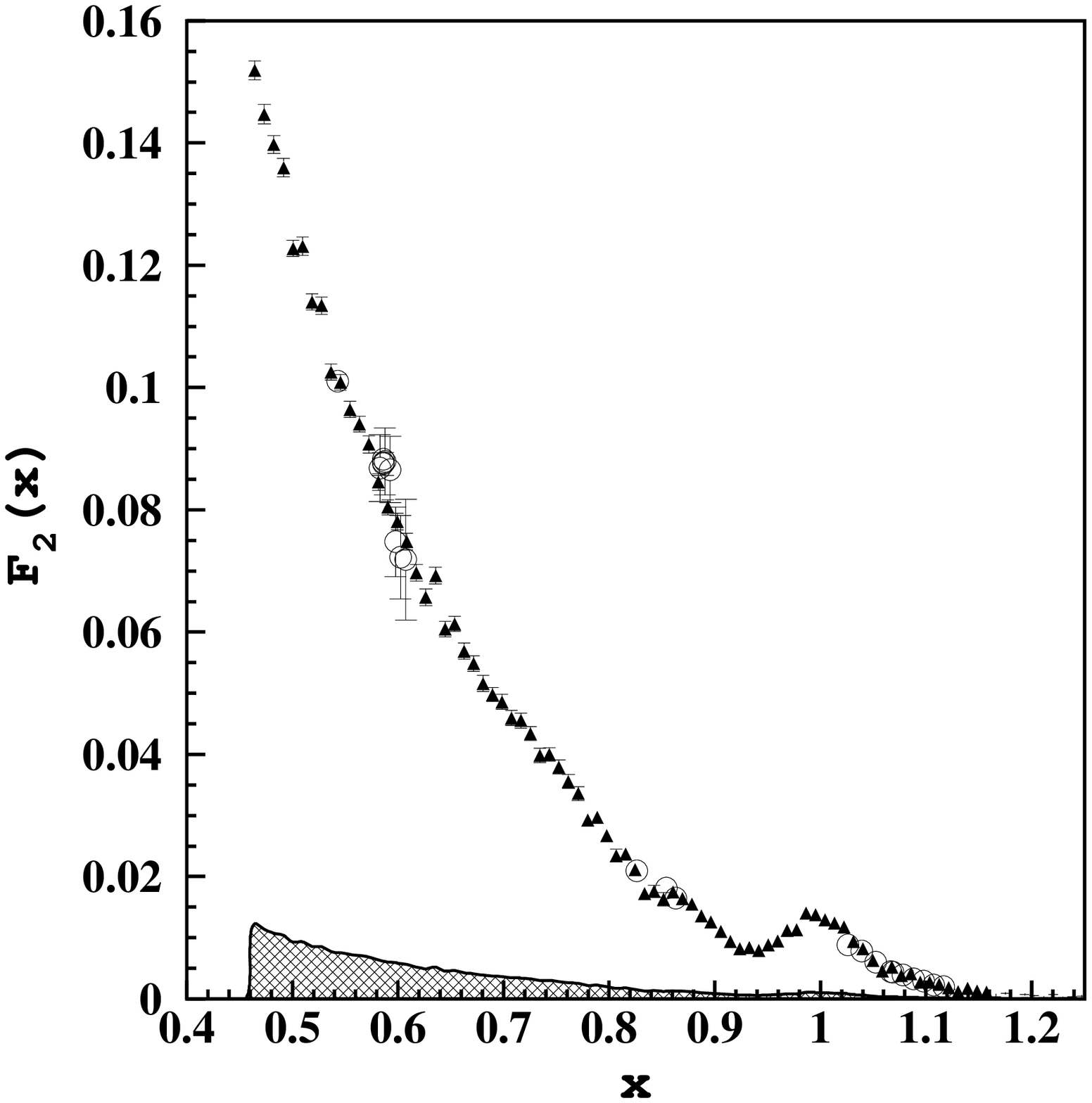}~%
\includegraphics[bb=1cm 5cm 19cm 23cm, scale=0.4]{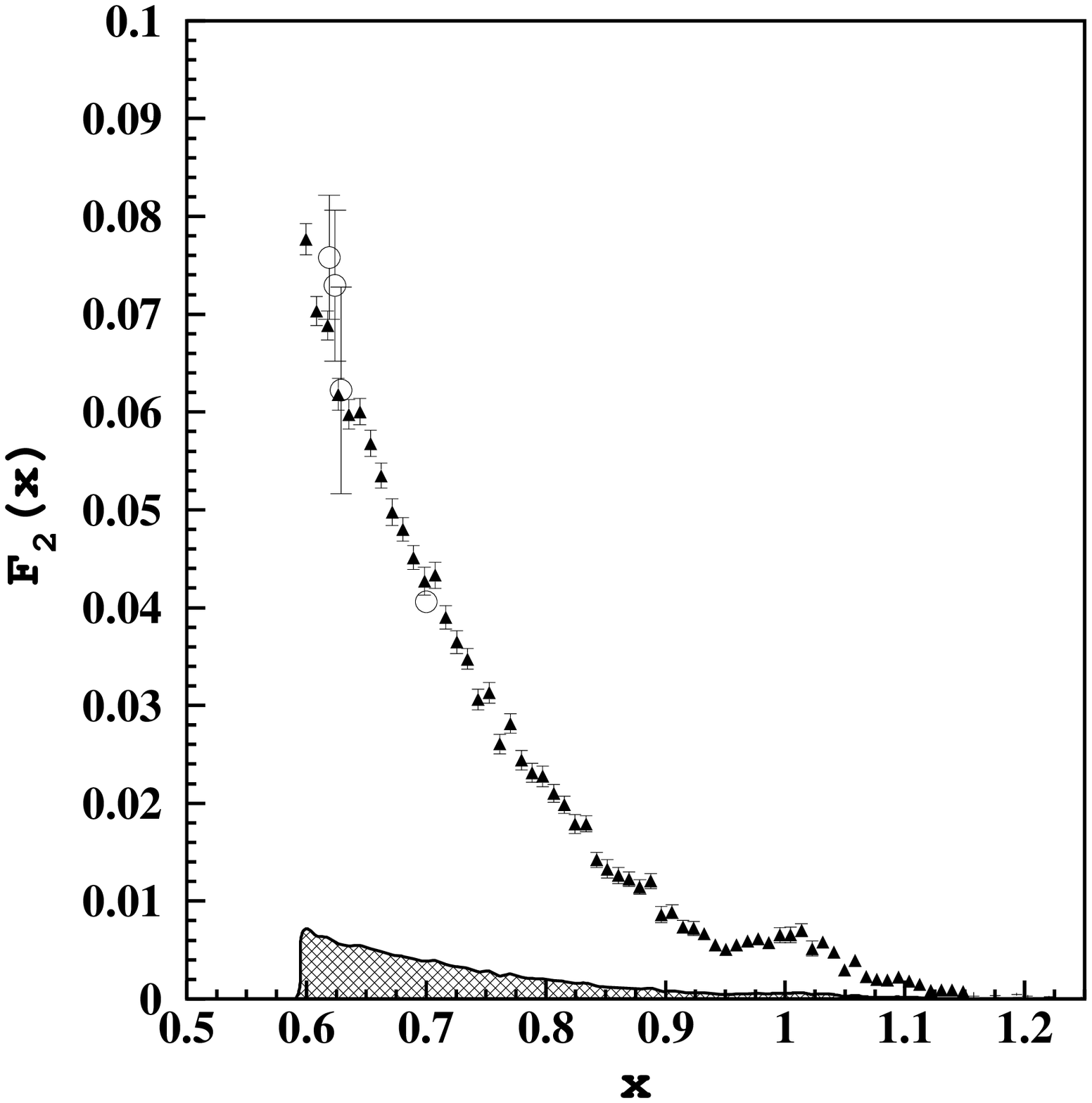}
\caption{\label{fig:f2comp} The deuteron structure function $F_2(x,Q^2)$
per nucleon at six different $Q^2$ values: from top-left in order
$Q^2=0.825$ (GeV/c)$^2$, $Q^2=1.375$ (GeV/c)$^2$, $Q^2=1.775$ (GeV/c)$^2$.
$Q^2=2.825$ (GeV/c)$^2$, $Q^2=4.075$ (GeV/c)$^2$ and $Q^2=5.025$ (GeV/c)$^2$.
The triangles represent experimental data obtained in the present analysis
with systematic uncertainties indicated by the hatched area.
The empty circles show data from previous experiments
~\cite{f2-hc,f2-hc_qe,E133,E140,E140x,E49a,E49b,E61,E891,E892,NE11}.}
\end{figure*}

Fig.~\ref{fig:f2comp} shows a comparison between the $F_2$ data
from CLAS and the other world data in a few $Q^2$ bins.
The CLAS data agree very well with all previous measurements.
The values of $F_2(x,Q^2)$, together with their statistical and systematic
uncertainties, are tabulated elsewhere~\cite{osipenko_f2d_CLAS_Note}.

In the calculation of the radiative correction factor $\Psi_{rad}$
we used the cross section model described in Appendix~\ref{app:f2_model}
in the following way:
\begin{itemize}
\item the $eD$ elastic radiative tail was calculated according to the ``exact''
Mo and Tsai formula~\cite{Mo};
\item in the quasi-elastic peak region ($W^{el}+\Delta W<W<1.2$ GeV) we applied
the correction formula to the continuum spectrum given in Ref.~\cite{Mo},
which is
based on the peaking approximation and is known to be reliable only
when $E^\prime/E > 0.5$. Here $W^{el}$ is the $eD$ elastic peak position
and $\Delta W$ its width;
\item at $W>1.2$ GeV we applied the exact Mo and Tsai formula also to the
quasi-elastic tail
and a peaking approximation based formula (referred to as the ``unfolding procedure'')
to the inelastic spectrum. For an exact calculation  of the
quasi-elastic tail
it was necessary to extract quasi-elastic form-factors. To this end
we integrated the quasi-elastic cross section from the beginning
of the peak up to $W=1.2$ GeV and performed a separation of the electric and magnetic
form-factors. These two kinematic regions overlap quite well
and don't exhibit any discontinuity at the point $W=1.2$ GeV. This
assured us that the peaking approximation formalism is safely applicable
to the quasi-elastic tail up to $W=1.2$ GeV.
\end{itemize}
The radiative correction factor $\Psi_{rad}$ varies strongly
in the explored kinematic range from 0.7 up to 1.5. Fortunately,
the largest corrections are given by the tails of the elastic and
quasi-elastic peaks,
for which calculations are very accurate (see Refs.~\cite{Mo,Akushevich}).
The largest systematic uncertainties (see Table~\ref{table:syserr})
are due to the efficiency evaluation, and to the photo-electron and
radiative corrections.

The systematic uncertainties for the efficiency evaluation, the
$e^+e^-$ pair correction, and the pion rejection correction
were described above, and the systematic uncertainties 
arising from the applied CLAS momentum correction routines
are calculated according to Ref.~\cite{osipenko_f2p}.
The radiative correction factors in $\sigma^M_{rad}$ were evaluated with two
different methods (\cite{Mo},~\cite{Akushevich}) and the difference was
taken as an estimate of the corresponding systematic uncertainty.
These two methods use different parameterizations of the elastic
(\cite{Stuart_ed} and \cite{Akushevich}), quasi-elastic (\cite{Simula_QE}
and \cite{Akushevich})
and inelastic (\cite{f2-hc} and~\cite{Akushevich}) cross sections, as well as
different calculation techniques.
The uncertainties in $R$ given in Appendix~\ref{app:rlt_model} were propagated
to the resulting $F_2$.
All systematic uncertainties were summed in quadrature to obtain the
final systematic uncertainty.

The statistical and systematic precisions of the extracted
structure function $F_2$ are strongly dependent on the kinematics:
the statistical uncertainties vary from 0.1\% up to 30\% at the largest $Q^2$,
where the event yield is very limited, while the average value is about 3\%;
the systematic uncertainties range from 4\% up to 14\%, with
the mean value being about 7\% (see Table~\ref{table:syserr}).

\begin{table}[!h]
\begin{center}
\caption{Range and average of systematic uncertainties on $F_2$.}
\label{table:syserr}
\vspace{2mm}
\begin{tabular}{|c|c|c|} \hline
     Source of uncertainties                 & Variation range   & Average \\ \cline{2-3}
                                             & [\%]              & [\%]    \\ \hline
Efficiency evaluation                        & 3-7               & 4.6  \\ \hline
$e^+e^-$ pair production correction          & 0-3               & 0.1  \\ \hline
Pion rejection correction                    & 0.1-6             & 2.4  \\ \hline
Radiative correction                         & 1.8-3.5           & 2.3  \\ \hline
Momentum correction                          & 0.2-1.2           & 0.5  \\ \hline
Uncertainty of $R=\frac{\sigma_L}{\sigma_T}$ & 0.2-0.75          & 0.5  \\ \hline
Empty target subtraction                     & 0.4-0.42          & 0.4  \\ \hline
Total                                        & 4-14              & 6.6  \\ \hline
\end{tabular}
\end{center}
\end{table}

\subsection{Moments of the Structure Function $F_2$}\label{sec:d_nm}
\begin{figure}
\includegraphics[bb=1cm 6cm 20cm 23cm, scale=0.4]{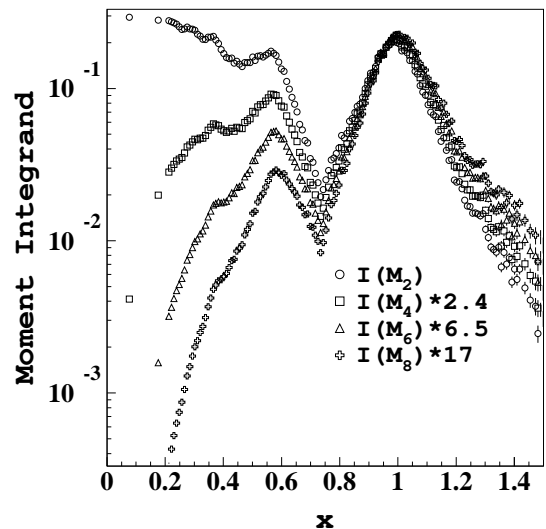}
\caption{\label{fig:intgr} Integrands of the Nachtmann moments at
$Q^2=0.825$ (GeV/c)$^2$: circles represent the integrand of the $M_2$;
squares show the integrand of the $M_4$;
triangles show the integrand of the $M_6$;
crosses show the integrand of the $M_8$.}
\end{figure}

\begin{figure}
\includegraphics[bb=1cm 6cm 20cm 23cm, scale=0.4]{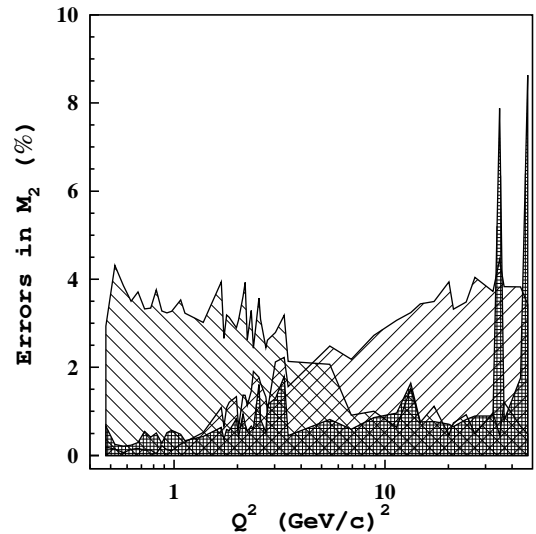}
\caption{\label{fig:SepErr} Uncertainties of the Nachtmann moment $M_2$
in percentage. The
lower cross-hatched area represents statistical uncertainties. The
left-hatched area represents the systematic uncertainties. The right-hatched
area represents the low-$x$ extrapolation uncertainty.}
\end{figure}

The evaluation of the deuteron structure function moments was performed
according to the method developed in Ref.~\cite{osipenko_f2p}.
However, there are two main differences in the deuteron data analysis:
\begin{enumerate}
\item the quasi-elastic peak is not as well known as the proton elastic
form-factors,
and moreover cannot be easily separated from the inelastic spectrum. Hence,
in contrast to the free proton target, we extract the total moments
of the deuteron structure function $F_2$ directly, without separating them
into the elastic and inelastic parts. This emphasizes the importance of a
precise determination
of the quasi-elastic peak for each $Q^2$ together with the inelastic spectrum.
In particular, the contribution of the quasi-elastic peak in the higher
moments ($n>2$)
at $Q^2$ values in the interval 1-5 (GeV/c)$^2$ is very significant,
as one can see in Fig.~\ref{fig:intgr}. Thanks to the CLAS data this problem
is well addressed now;
\item the lack of collider data does not allow one to reach very low $x$
values.
For the second moment $M_2$ this leads to an increase of systematic
uncertainties
due to the low-$x$ extrapolation with respect to the proton $M_2$.
The contribution of the low $x$ part in the higher moments however is
negligible. We used two models of the deuteron structure
function $F_2$ which have very different low-$x$ behaviour to estimate
the extrapolated part of the second moment $M_2$. The difference between
the two estimates
was taken as an evaluation of the corresponding systematic uncertainty.
A comparison
of the extrapolation systematic uncertainties and other uncertainties in $M_2$
is shown in Fig.~\ref{fig:SepErr}.
\end{enumerate}

We combined the structure functions $F_2$ obtained from the CLAS data and
the other world data on the structure function $F_2$, along with the inclusive
cross section data
from Refs.~\cite{f2-hc,f2-hc_qe,E133,E140,E140x,E49a,E49b,E61,E891,E892,NE11,BCDMS,NMC,NMC_f2d_f2p_ratio,E665,E665_f2d_f2p_ratio}
(see Fig.~\ref{fig:xandQ2Domain}).
The data from Ref.~\cite{BEBC}, recently reanalyzed in Ref.~\cite{Barone}
with the inclusion of radiative and bin centering corrections,
are not used in the present analysis due to large statistical uncertainties and
unknown systematic uncertainties.
The $Q^2$-range of the CLAS data, from 0.4 to 5.95 (GeV/c)$^2$,
was divided
into bins of width $\Delta Q^2 =$ 0.05~(GeV/c)$^2$. Within each
$Q^2$ bin the world data were
shifted to the central bin value $Q^2_0$, using the fit of $F_2^M(x,Q^2)$ from
Appendix.~\ref{app:f2_model}.
The integrals of the data over $x$ were performed numerically using the
standard trapezoidal method TRAPER~\cite{cernlib}.
As an example, Fig.~\ref{fig:intgr} shows the integrands of the first four
moments
as a function of $x$ at fixed $Q^2$. The significance of the large $x$
region for various moments can clearly be seen.

\begin{figure}
\includegraphics[bb=1cm 6cm 20cm 23cm, scale=0.4]{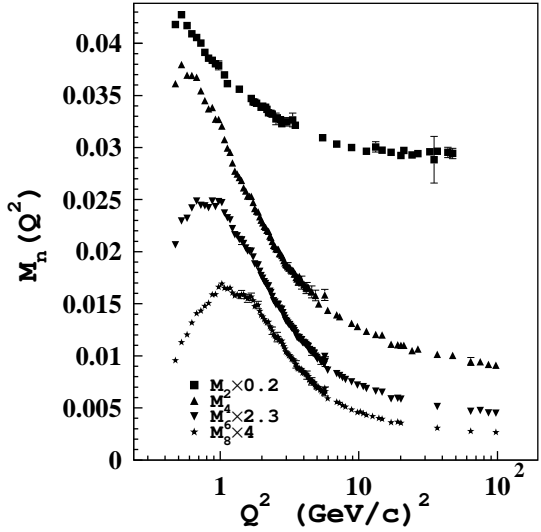}
\caption{\label{fig:NachtMom}
The Nachtman moments extracted from the world data,
including the new CLAS results. Uncertainties are statistical only.}
\end{figure}

As in Ref.~\cite{osipenko_f2p}, the world data at $Q^2$ above
$6$~(GeV/c)$^2$ were analyzed in the same way as described above,
but with a different $Q^2$ bin size. The bin size was chosen for the
data to provide sufficient $x$-coverage for most of the
$Q^2$ bins ($\Delta Q^2/Q^2 =5$ \%).
The results together with their statistical and
systematic uncertainties are shown in Fig.~\ref{fig:NachtMom}
and reported in Table~\ref{table:r_nm1}.

The systematic uncertainty consists of experimental uncertainties in the
data given in Refs.~\cite{f2-hc,f2-hc_qe,E133,E140,E140x,E49a,E49b,E61,E891,E892,NE11,BCDMS,NMC,NMC_f2d_f2p_ratio,E665,E665_f2d_f2p_ratio}
and uncertainties in the evaluation procedure. To estimate the first type
of uncertainty we had to account for the inclusion of many data sets
measured in different laboratories with different detectors.
In the present analysis we assume that the
different experiments are independent and therefore
only the systematic uncertainties within a given data set are correlated.

\section{Separation of Leading and higher twists}\label{sec:Discussion}
In order to separate the leading and higher twists in the measured moments
we used the method developed in Refs.~\cite{Ricco1,SIM00,osipenko_f2p}.
This method is essentially based on a general form of the OPE for
the structure function moments, where the leading twist $Q^2$-evolution
is calculated in pQCD and the deviation from this behaviour is assigned
to the higher twist contribution. Therefore the measured Nachtmann
$n$-th moment was parameterized as:
\begin{eqnarray}
     M_n^N(Q^2) = \eta_n(Q^2) + HT_n(Q^2) ~ ,
    \label{eq:twists}
\end{eqnarray}
\noindent where $\eta_n(Q^2)$ is the leading twist moment and
$HT_n(Q^2)$ is the higher twist contribution. The leading twist term
was calculated including soft gluon re-summation (SGR) corrections which go beyond
the fixed-order next-to-leading approximation and are essential for
a reliable extraction of higher twists~\cite{SIM00}.
The observed decoupling of the singlet quark and gluon densities
at large $x$~\cite{Ricco1} allows us to consider only the non-singlet (NS)
evolution for $n \geq 4$ and therefore to reduce the number of leading twist parameters.
In this approximation the leading twist moment $\eta_n(Q^2)$ for $n \geq 4$
can be written as follows:
 \begin{eqnarray}
       \eta_n(Q^2) & = & A_n \left[ {\alpha_s(Q^2) \over 
       \alpha_s(\mu^2)} \right]^{\gamma_n^{NS}} \nonumber \\
       & & \left\{ \left[1 +  {\alpha_s(Q^2) \over 2 \pi}
       C_{DIS}^{(NLO)} \right]  e^{G_n(Q^2)} + 
       \right.  \nonumber \\
       & & \left. {\alpha_s(Q^2) \over 4 \pi} R_n^{NS} \right\} ~ , 
       \label{eq:SGR}
 \end{eqnarray}
\noindent where the quantities $\gamma_n^{NS}$, $C_{DIS}^{(NLO)}$
and $R_n^{NS} $ can be obtained from Ref.~\cite{SIM00},
$\alpha_s(M_Z^2) = 0.118$~\cite{PDG} and
the reference scale $\mu^2 = 10$~(GeV/c)$^2$.
In Eq.~\ref{eq:SGR} the function $G_n(Q^2)$ is the key quantity of
the soft gluon re-summation. At Next-to-Leading-Log it reads as:
\begin{eqnarray}
       G_n(Q^2) = \mbox{ln}(n) G_1(\lambda_n) + G_2(\lambda_n) +O[\alpha_s^k 
       \mbox{ln}^{k-1}(n)] ~ ,
       \label{eq:Gn}
\end{eqnarray}
where $\lambda_n \equiv \beta_0 \alpha_s(Q^2) \mbox{ln}(n) / 4\pi$ and:
 \begin{eqnarray}
       G_1(\lambda) & = & C_F {4 \over \beta_0 \lambda} \left[ \lambda + (1 
       -  \lambda) \mbox{ln}(1 - \lambda) \right] ~ , \nonumber \\
       G_2(\lambda) & = & - C_F {4 \gamma_E + 3 \over \beta_0} \mbox{ln}(1 
       - \lambda) \nonumber \\
       & & - C_F {8 K \over \beta_0^2} \left[ \lambda + \mbox{ln}(1 
       - \lambda) \right] \nonumber \\
      & & + C_F {4  \beta_1 \over \beta_0^3} \left[ \lambda + \mbox{ln}(1 - 
      \lambda) + {1 \over 2} \mbox{ln}^2(1 - \lambda) \right] ~,~~~~~~ 
     \label{eq:G1G2}
 \end{eqnarray}
\noindent with $C_F \equiv (N_c^2-1)/(2N_c)$,
$k = N_c (67/18 - \pi^2 / 6) - 5 N_f / 9$,
$\beta_0 = 11 - 2 N_f / 3$, and $N_f$ being the number of active
flavors. Note that the function $G_2(\lambda)$ is divergent for
$\lambda \to 1$. This means that at large $n$ (i.e. large $x$)
the soft gluon re-summation cannot be extended to arbitrarily low
values of $Q^2$. Therefore, for a safe use of present SGR
techniques we work far from the above-mentioned divergences
by limiting our analysis of low-order moments ($n \leq 8$) to
$Q^2 \geq 0.7-1$~(GeV/c)$^2$.

\begin{figure*}
\includegraphics[bb=1cm 4cm 20cm 23cm, scale=0.4]{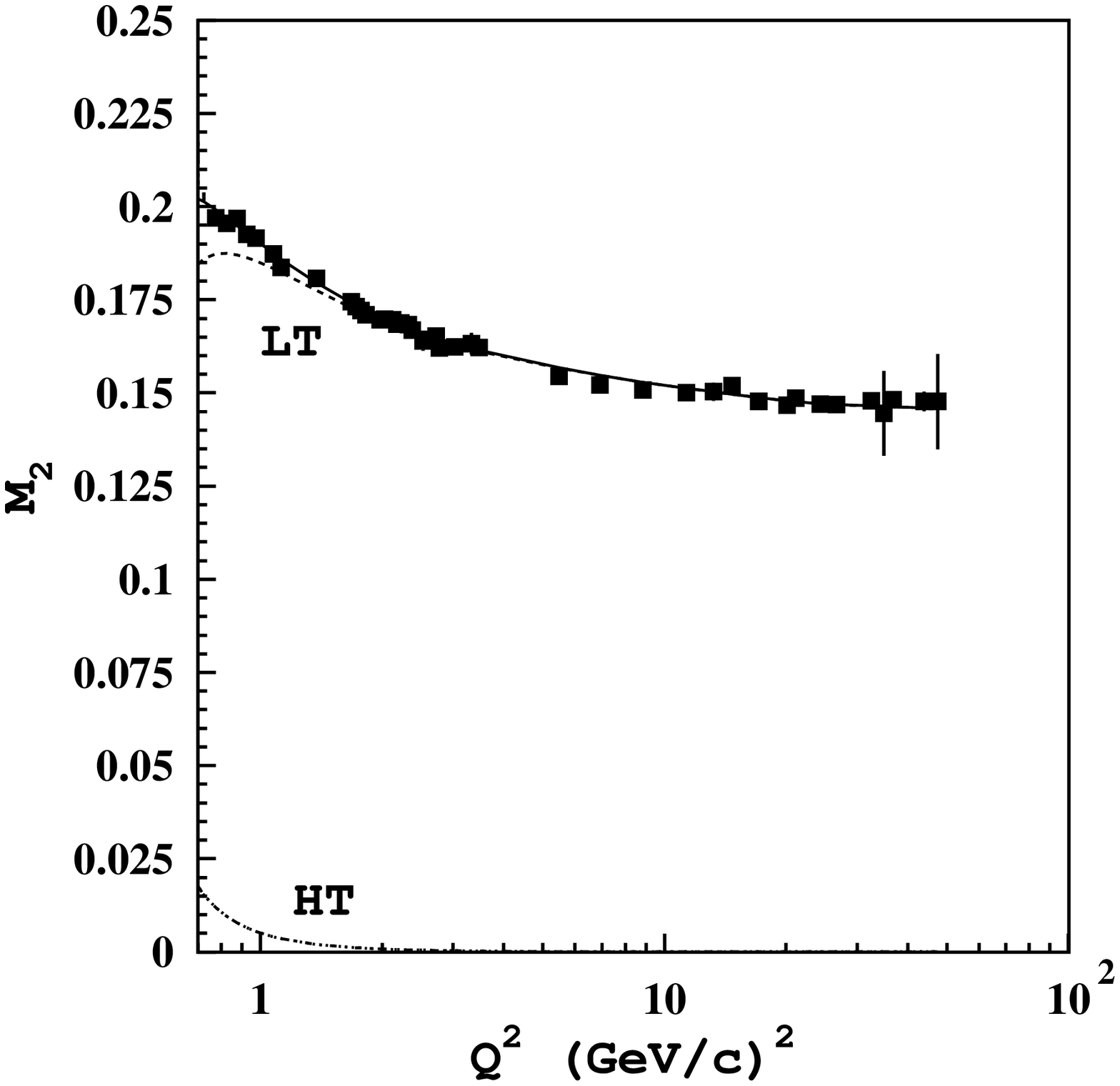}~~~~%
\includegraphics[bb=1cm 4cm 20cm 23cm, scale=0.4]{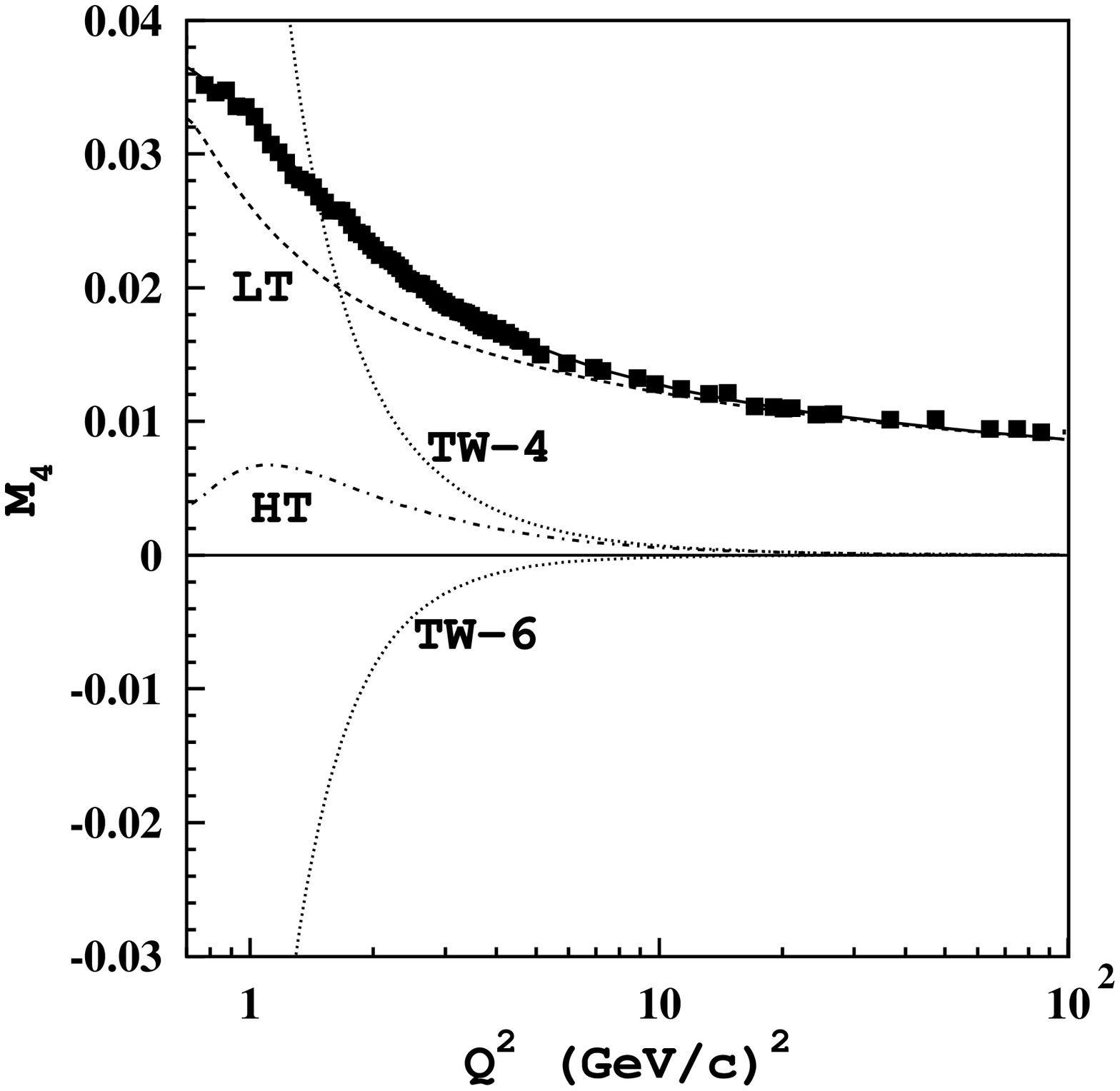}
\includegraphics[bb=1cm 4cm 20cm 23cm, scale=0.4]{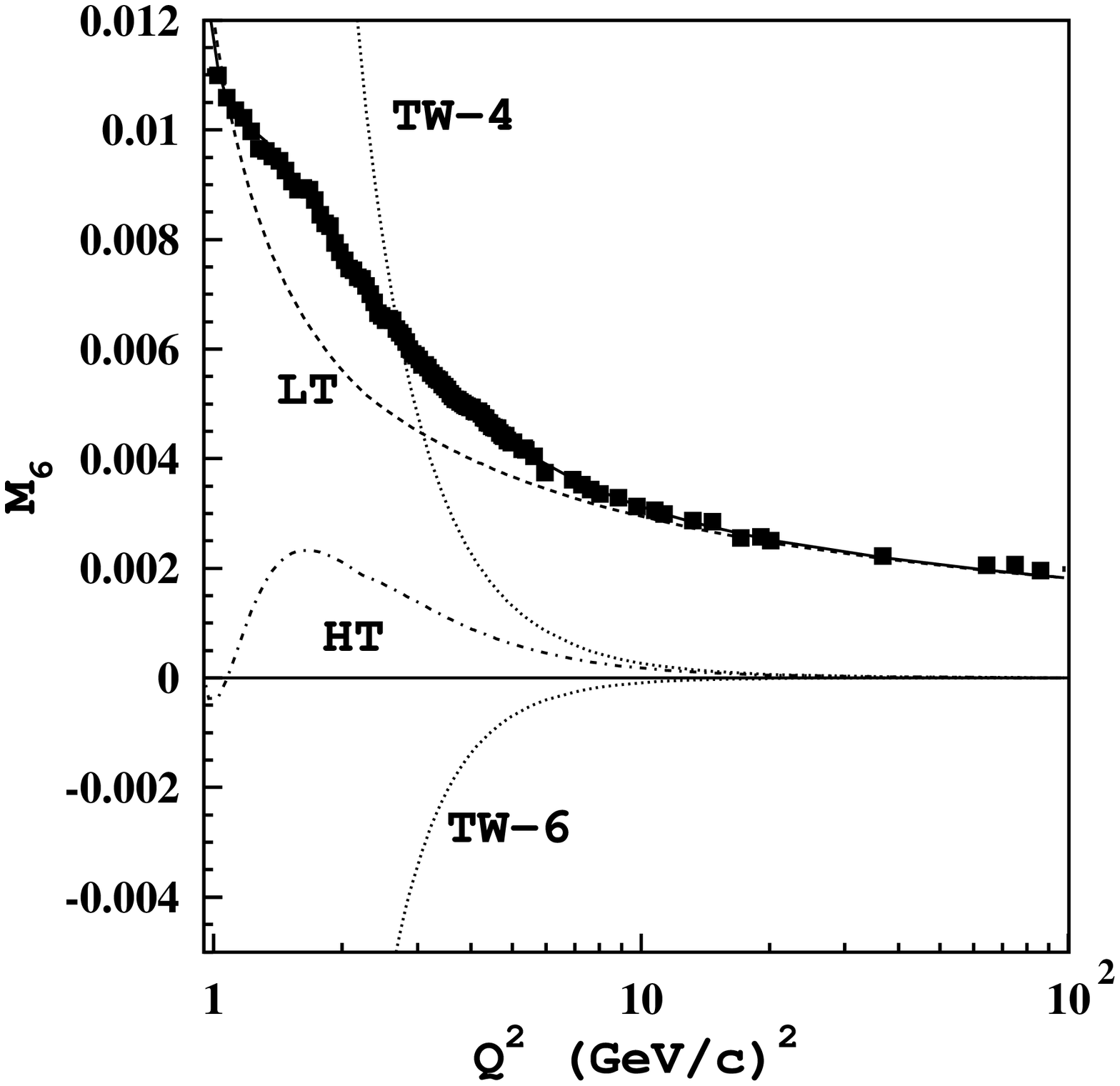}~~~~%
\includegraphics[bb=1cm 4cm 20cm 23cm, scale=0.4]{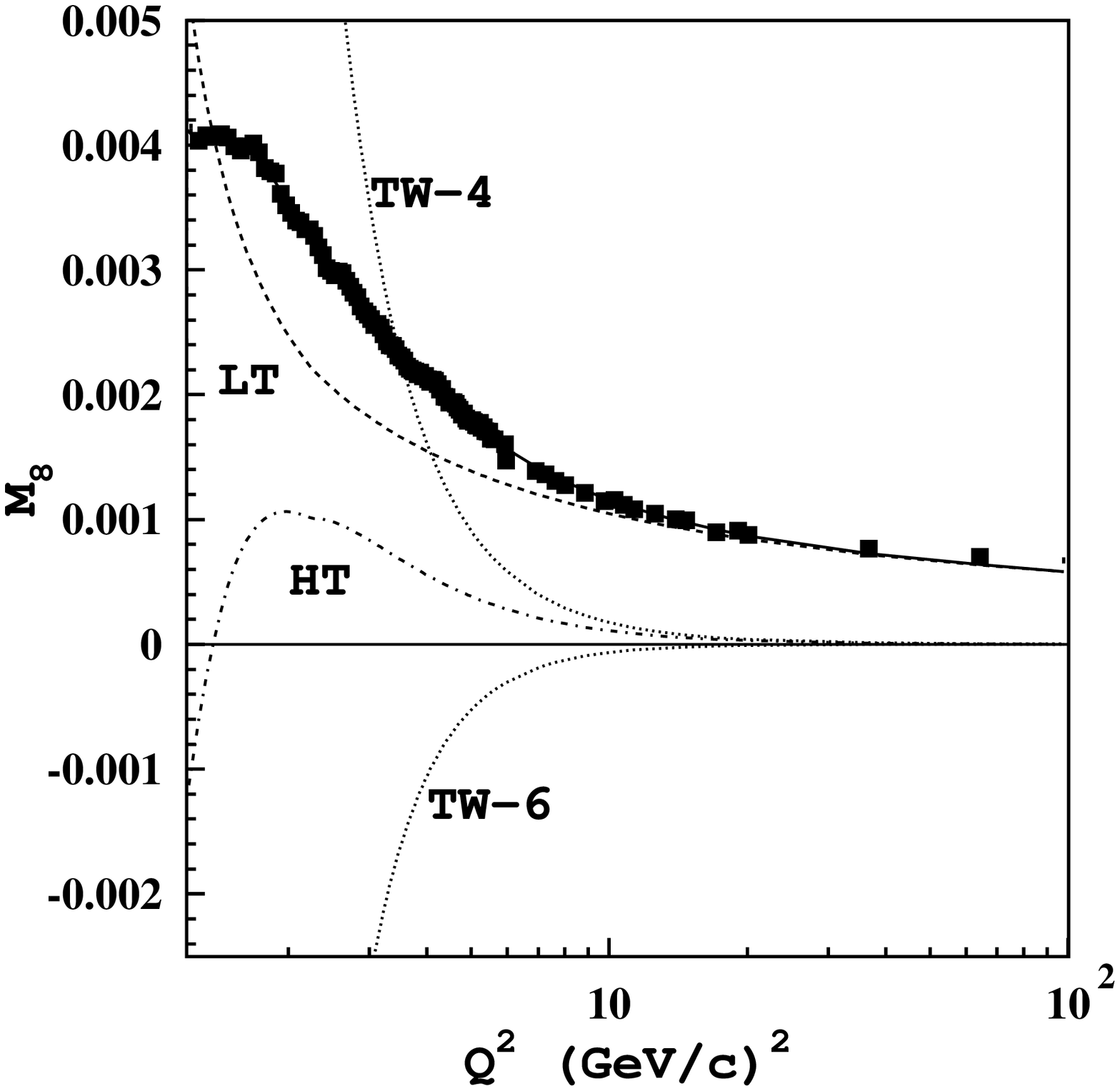}
\caption{\label{fig:twists}
Results of the twist analysis.  The squares
represent the Nachtman moments obtained in this analysis.  The
solid line is the fit to the moments using Eq.~\ref{eq:twists} with the
parameters listed in Table~\ref{table:twist1}.  The twist-2, twist-4, twist-6
and higher twist (HT) contributions to the fit are indicated.
The twist-2 contribution was calculated using Eq.~\ref{eq:SGR}.}
\end{figure*}

Since a complete calculation of the higher twist anomalous dimensions is
not yet available, we use the same phenomenological ansatz already adopted
in Refs.~\cite{Ricco1,SIM00,osipenko_f2p}.
In this approach the higher twist contribution is given by~\cite{Ji}:
\begin{eqnarray}\nonumber
HT_n(Q^2)=&&
a_n^{(4)}\biggl[\frac{\alpha_s(Q^2)}{\alpha_s(\mu^2)}\biggr]^{
\gamma_n^{(4)}}\frac{\mu^2}{Q^2}\\
+&&a_n^{(6)}\biggl[\frac{\alpha_s(Q^2)}{\alpha_s(\mu^2)}\biggr]^{
\gamma_n^{(6)}}\frac{\mu^4}{Q^4} ~ ,
\label{eq:HT}
\end{eqnarray}
\noindent where the logarithmic pQCD evolution of the twist-$\tau$
contribution is accounted for by the term
$[\alpha_s(Q^2) / \alpha_s(\mu^2)]^{\gamma_n^{(\tau)}}$.
This term corresponds to the Wilson coefficient $E_{n \tau}(\mu_r,\mu_f,Q^2)$
in Eq.~\ref{eq:i_m1} with an {\em effective} anomalous dimension
$\gamma_n^{(\tau)}$. The parameter $a_n^{(\tau)}$ represents
the overall strength of the twist-$\tau$ term at the
renormalization scale $Q^2 = \mu^2$ and it is proportional to the
matrix element $O_{n \tau}(\mu)$ in Eq.~\ref{eq:i_m1}.
The presence of two distinct higher twist terms for $n \ge 4$ is motivated
by the $Q^2$-behaviour of the total higher twist contribution
in the moments. This was obtained by a direct subtraction of the leading twist
term fitted to the large $Q^2$ part of the plot from the measured moments.
Existence of maxima and moreover of the sign turn-over
(see Fig.~\ref{fig:twists})
in the total higher twist contribution cannot be described by a single
twist term
within the pQCD-inspired model from Eq.~\ref{eq:HT}.
Therefore, the presence of at least two higher twist terms is necessary
for a successful description of experimental data.
We checked that the variation of the total higher twist contribution
after inclusion of twist-8 and twist-10 terms is smaller than
the quoted systematic uncertainties.

The $n$-th moment (see Eqs.~\ref{eq:twists}, \ref{eq:SGR} and \ref{eq:HT})
for $n \ge 4$ has five unknown parameters:
the twist-2 parameter $A_n$ and the higher twist parameters
$a_n^{(4)}, \gamma_n^{(4)}, a_n^{(6)}, \gamma_n^{(6)}$.
All five unknown parameters were simultaneously determined from
a $\chi^2$-minimization procedure in the $Q^2$ range between 1 and
100~(GeV/c)$^2$.
In this procedure only the statistical uncertainties of the experimental
moments
were taken into account. The uncertainties of the various twist parameters
were then obtained by adding the systematic uncertainties to the experimental
moments and by repeating the twist extraction procedure. 

For $n=2$, $\eta_2(Q^2)$ is given by the sum of the non-singlet
and singlet terms, which yield two unknown parameters associated
with the leading twist. These parameters are the values of the gluon
and non-singlet quark moments at the reference scale $Q^2 = \mu^2$.
However, due to the vanishing contribution of the higher twists in $M_2$
(see Fig.~\ref{fig:twists})
one can reduce the number of parameters in $HT_n(Q^2)$
by limiting the expansion to the twist-4 term only.

The parameter values obtained at the reference scale
$\mu^2=10$ (GeV/c)$^2$ are reported
in Table~\ref{table:twist1}, where it can be seen
that the leading twist is determined with an uncertainty of a few percent,
while the precision of the extracted higher twists decreases with $n$,
reaching an overall 20-30\% for $n=8$,
thanks to the CLAS data at large $x$.
Note that the leading twist is
directly extracted from the data, which means that no specific
functional shape of the parton distributions is assumed in
our analysis.

\begin{table*}
\caption{\label{table:r_nm1}The Nachtmann moments for
$n=2,4,6$ and $8$ evaluated in the interval
$0.05 \le~Q^2 \le 100$~(GeV/c)$^2$. The moments are labeled
with an asterisk when the contribution to the integral by the experimental
data is
between 50\% and 70\%. All the others were evaluated
with more than 70\% data coverage. The data are reported
together with the statistical and systematic uncertainties,
the third uncertainty for $n=2$ is due to low-$x$
extrapolation.}
\begin{ruledtabular}
\begin{tabular}{|c|c|c|c|c|} \cline{1-5}
$Q^2~[$(GeV/c)$^2]$ & $M_2(Q^2)$x$10^{-1}$ & $M_4(Q^2)$x$10^{-2}$ & $M_6(Q^2)$x$10^{-2}$ & $M_8(Q^2)$x$10^{-3}$ \\ \cline{1-5}
 0.475 & 2.133 $\pm$ 0.014 $\pm$ 0.063 $\pm$ 0.004 & 3.673 $\pm$ 0.035 $\pm$ 0.139 & 0.911 $\pm$ 0.010 $\pm$ 0.037 & 2.421 $\pm$ 0.030 $\pm$ 0.101 \\ \cline{1-5}
 0.525 & 2.168 $\pm$ 0.005 $\pm$ 0.093 $\pm$ 0.003 & 3.868 $\pm$ 0.017 $\pm$ 0.171 & 1.011 $\pm$ 0.005 $\pm$ 0.046 & 2.852 $\pm$ 0.016 $\pm$ 0.129 \\ \cline{1-5}
 0.575 & 2.118 $\pm$ 0.004 $\pm$ 0.082 $\pm$ 0.001 & 3.776 $\pm$ 0.011 $\pm$ 0.168 & 1.026 $\pm$ 0.004 $\pm$ 0.047 & 3.042 $\pm$ 0.012 $\pm$ 0.140 \\ \cline{1-5}
 0.625 & 2.084 $\pm$ 0.005 $\pm$ 0.073 $\pm$ 0.003 & 3.757 $\pm$ 0.008 $\pm$ 0.162 & 1.067 $\pm$ 0.003 $\pm$ 0.048 & 3.328 $\pm$ 0.010 $\pm$ 0.154 \\ \cline{1-5}
 0.675 & 2.064 $\pm$ 0.006 $\pm$ 0.077 $\pm$ 0.003 & 3.757 $\pm$ 0.009 $\pm$ 0.179 & 1.101 $\pm$ 0.003 $\pm$ 0.053 & 3.572 $\pm$ 0.013 $\pm$ 0.172 \\ \cline{1-5}
 0.725 & 2.027 $\pm$ 0.011 $\pm$ 0.067 $\pm$ 0.002 & 3.632 $\pm$ 0.010 $\pm$ 0.168 & 1.086 $\pm$ 0.004 $\pm$ 0.053 & 3.641 $\pm$ 0.016 $\pm$ 0.178 \\ \cline{1-5}
 0.775 & 1.970 $\pm$ 0.008 $\pm$ 0.066 $\pm$ 0.003 & 3.515 $\pm$ 0.012 $\pm$ 0.155 & 1.080 $\pm$ 0.005 $\pm$ 0.051 & 3.752 $\pm$ 0.022 $\pm$ 0.182 \\ \cline{1-5}
 0.825 & 1.955 $\pm$ 0.010 $\pm$ 0.073 $\pm$ 0.001 & 3.460 $\pm$ 0.012 $\pm$ 0.174 & 1.079 $\pm$ 0.005 $\pm$ 0.055 & 3.840 $\pm$ 0.024 $\pm$ 0.195 \\ \cline{1-5}
 0.875 & 1.968 $\pm$ 0.005 $\pm$ 0.064 $\pm$ 0.004 & 3.474 $\pm$ 0.007 $\pm$ 0.175 & 1.108 $\pm$ 0.003 $\pm$ 0.058 & 4.056 $\pm$ 0.015 $\pm$ 0.213 \\ \cline{1-5}
 0.925 & 1.926 $\pm$ 0.010 $\pm$ 0.062 $\pm$ 0.003 & 3.357 $\pm$ 0.014 $\pm$ 0.166 & 1.082 $\pm$ 0.007 $\pm$ 0.057 & 4.042 $\pm$ 0.035 $\pm$ 0.213 \\ \cline{1-5}
 0.975 & 1.916 $\pm$ 0.011 $\pm$ 0.063 $\pm$ 0.002 & 3.352 $\pm$ 0.013 $\pm$ 0.162 & 1.106 $\pm$ 0.006 $\pm$ 0.058 & 4.238 $\pm$ 0.032 $\pm$ 0.225 \\ \cline{1-5}
 1.025 & 			                   & 3.280 $\pm$ 0.013 $\pm$ 0.169 & 1.099 $\pm$ 0.006 $\pm$ 0.060 & 4.320 $\pm$ 0.034 $\pm$ 0.239 \\ \cline{1-5}
 1.075 & 1.873 $\pm$ 0.009 $\pm$ 0.066 $\pm$ 0.005 & 3.159 $\pm$ 0.014 $\pm$ 0.164 & 1.059 $\pm$ 0.007 $\pm$ 0.059 & 4.215 $\pm$ 0.039 $\pm$ 0.240 \\ \cline{1-5}
 1.125 & 1.837 $\pm$ 0.006 $\pm$ 0.059 $\pm$ 0.006 & 3.071 $\pm$ 0.007 $\pm$ 0.147 & 1.036 $\pm$ 0.003 $\pm$ 0.056 & 4.190 $\pm$ 0.018 $\pm$ 0.235 \\ \cline{1-5}
 1.175 &  			                   & 3.012 $\pm$ 0.008 $\pm$ 0.132 & 1.022 $\pm$ 0.003 $\pm$ 0.052 & 4.201 $\pm$ 0.017 $\pm$ 0.228 \\ \cline{1-5}
 1.225 &  			                   & 2.934 $\pm$ 0.016 $\pm$ 0.148 & 0.997 $\pm$ 0.007 $\pm$ 0.055 & 4.136 $\pm$ 0.037 $\pm$ 0.234 \\ \cline{1-5}
 1.275 &  			                   & 2.842 $\pm$ 0.013 $\pm$ 0.147 & 0.965 $\pm$ 0.006 $\pm$ 0.053 & 4.035 $\pm$ 0.032 $\pm$ 0.230 \\ \cline{1-5}
 1.325 &  			                   & 2.809 $\pm$ 0.012 $\pm$ 0.154 & 0.961 $\pm$ 0.005 $\pm$ 0.055 & 4.080 $\pm$ 0.027 $\pm$ 0.238 \\ \cline{1-5}
 1.375 & 1.807 $\pm$ 0.008 $\pm$ 0.055 $\pm$ 0.009 & 2.789 $\pm$ 0.014 $\pm$ 0.138 & 0.951 $\pm$ 0.008 $\pm$ 0.053 & 4.064 $\pm$ 0.046 $\pm$ 0.238 \\ \cline{1-5}
 1.425 &  			                   & 2.751 $\pm$ 0.011 $\pm$ 0.130 & 0.944 $\pm$ 0.006 $\pm$ 0.051 & 4.086 $\pm$ 0.033 $\pm$ 0.235 \\ \cline{1-5}
 1.475 &  			                   & 2.681 $\pm$ 0.016 $\pm$ 0.119 & 0.926 $\pm$ 0.007 $\pm$ 0.047 & 4.063 $\pm$ 0.041 $\pm$ 0.221 \\ \cline{1-5}
 1.525 &  			                   & 2.636 $\pm$ 0.015 $\pm$ 0.149 & 0.906 $\pm$ 0.007 $\pm$ 0.052 & 3.993 $\pm$ 0.041 $\pm$ 0.232 \\ \cline{1-5}
 1.575 &  			                   & 2.574 $\pm$ 0.014 $\pm$ 0.153 & 0.891 $\pm$ 0.007 $\pm$ 0.056 & 3.956 $\pm$ 0.043 $\pm$ 0.252 \\ \cline{1-5}
 1.625 &  			                   & 2.582 $\pm$ 0.011 $\pm$ 0.140 & 0.894 $\pm$ 0.005 $\pm$ 0.052 & 3.992 $\pm$ 0.025 $\pm$ 0.236 \\ \cline{1-5}
 1.675 & 1.745 $\pm$ 0.011 $\pm$ 0.069 $\pm$ 0.019 & 2.575 $\pm$ 0.013 $\pm$ 0.135 & 0.891 $\pm$ 0.007 $\pm$ 0.051 & 4.012 $\pm$ 0.042 $\pm$ 0.239 \\ \cline{1-5}
 1.725 & 1.733 $\pm$ 0.006 $\pm$ 0.046 $\pm$ 0.011 & 2.527 $\pm$ 0.008 $\pm$ 0.115 & 0.872 $\pm$ 0.005 $\pm$ 0.046 & 3.942 $\pm$ 0.033 $\pm$ 0.225 \\ \cline{1-5}
 1.775 & 1.721 $\pm$ 0.010 $\pm$ 0.055 $\pm$ 0.019 & 2.468 $\pm$ 0.009 $\pm$ 0.109 & 0.845 $\pm$ 0.006 $\pm$ 0.041 & 3.816 $\pm$ 0.035 $\pm$ 0.200 \\ \cline{1-5}
 1.825 & 1.709 $\pm$ 0.010 $\pm$ 0.054 $\pm$ 0.021 & 2.413 $\pm$ 0.010 $\pm$ 0.111 & 0.829 $\pm$ 0.007 $\pm$ 0.041 & 3.789 $\pm$ 0.048 $\pm$ 0.193 \\ \cline{1-5}
 1.875 &  			                   & 2.402 $\pm$ 0.013 $\pm$ 0.117 & 0.824 $\pm$ 0.006 $\pm$ 0.045 & 3.774 $\pm$ 0.031 $\pm$ 0.212 \\ \cline{1-5}
 1.925 &  			                   & 2.349 $\pm$ 0.018 $\pm$ 0.112 & 0.794 $\pm$ 0.008 $\pm$ 0.041 & 3.610 $\pm$ 0.047 $\pm$ 0.200 \\ \cline{1-5}
 1.975 & 1.697 $\pm$ 0.014 $\pm$ 0.049 $\pm$ 0.023 & 2.315 $\pm$ 0.012 $\pm$ 0.105 & 0.777 $\pm$ 0.008 $\pm$ 0.039 & 3.519 $\pm$ 0.060 $\pm$ 0.186 \\ \cline{1-5}
 2.025 & 1.698 $\pm$ 0.007 $\pm$ 0.052 $\pm$ 0.011 & 2.280 $\pm$ 0.005 $\pm$ 0.106 & 0.762 $\pm$ 0.003 $\pm$ 0.038 & 3.456 $\pm$ 0.018 $\pm$ 0.178 \\ \cline{1-5}
 2.075 &  			                   & 2.246 $\pm$ 0.011 $\pm$ 0.104 & 0.747 $\pm$ 0.004 $\pm$ 0.037 & 3.395 $\pm$ 0.023 $\pm$ 0.171 \\ \cline{1-5}
 2.125 & 1.696 $\pm$ 0.016 $\pm$ 0.061 $\pm$ 0.023 & 2.244 $\pm$ 0.010 $\pm$ 0.110 & 0.744 $\pm$ 0.007 $\pm$ 0.038 & 3.383 $\pm$ 0.054 $\pm$ 0.172 \\ \cline{1-5}
 2.175 & 1.684 $\pm$ 0.012 $\pm$ 0.066 $\pm$ 0.023 & 2.211 $\pm$ 0.010 $\pm$ 0.109 & 0.731 $\pm$ 0.007 $\pm$ 0.037 & 3.324 $\pm$ 0.052 $\pm$ 0.172 \\ \cline{1-5}
 2.225 & 1.687 $\pm$ 0.009 $\pm$ 0.044 $\pm$ 0.019 & 2.201 $\pm$ 0.004 $\pm$ 0.103 & 0.728 $\pm$ 0.002 $\pm$ 0.036 & 3.324 $\pm$ 0.016 $\pm$ 0.168 \\ \cline{1-5}
 2.275 &  			                   & 2.168 $\pm$ 0.019 $\pm$ 0.099 & 0.716 $\pm$ 0.004 $\pm$ 0.035 & 3.274 $\pm$ 0.023 $\pm$ 0.162 \\ \cline{1-5}
 2.325 & 1.683 $\pm$ 0.011 $\pm$ 0.055 $\pm$ 0.025 & 2.148 $\pm$ 0.006 $\pm$ 0.094 & 0.700 $\pm$ 0.004 $\pm$ 0.032 & 3.179 $\pm$ 0.029 $\pm$ 0.149 \\ \cline{1-5}
 2.375 & 1.669 $\pm$ 0.010 $\pm$ 0.041 $\pm$ 0.032 & 2.104 $\pm$ 0.005 $\pm$ 0.085 & 0.685 $\pm$ 0.003 $\pm$ 0.030 & 3.121 $\pm$ 0.027 $\pm$ 0.143 \\ \cline{1-5}
 2.425 &  			                   & 2.063 $\pm$ 0.006 $\pm$ 0.088 & 0.666 $\pm$ 0.003 $\pm$ 0.028 & 3.016 $\pm$ 0.021 $\pm$ 0.130 \\ \cline{1-5}
 2.475 &  			                   & 2.054 $\pm$ 0.008 $\pm$ 0.079 & 0.661 $\pm$ 0.003 $\pm$ 0.027 & 2.991 $\pm$ 0.020 $\pm$ 0.125 \\ \cline{1-5}
 2.525 & 1.639 $\pm$ 0.026 $\pm$ 0.059 $\pm$ 0.029 & 2.034 $\pm$ 0.005 $\pm$ 0.089 & 0.653 $\pm$ 0.002 $\pm$ 0.028 & 2.956 $\pm$ 0.019 $\pm$ 0.121 \\ \cline{1-5}
 2.575 & 1.644 $\pm$ 0.013 $\pm$ 0.051 $\pm$ 0.027 & 2.032 $\pm$ 0.008 $\pm$ 0.091 & 0.656 $\pm$ 0.005 $\pm$ 0.032 & 2.990 $\pm$ 0.042 $\pm$ 0.144 \\ \cline{1-5}
 2.625 &  			                   & 2.026 $\pm$ 0.009 $\pm$ 0.101 & 0.652 $\pm$ 0.004 $\pm$ 0.034 & 2.974 $\pm$ 0.032 $\pm$ 0.153 \\ \cline{1-5}
 2.675 &  			                   & 1.986 $\pm$ 0.008 $\pm$ 0.090 & 0.637 $\pm$ 0.003 $\pm$ 0.032 & 2.914 $\pm$ 0.021 $\pm$ 0.154 \\ \cline{1-5}
 2.725 & 1.652 $\pm$ 0.009 $\pm$ 0.040 $\pm$ 0.022 & 1.990 $\pm$ 0.004 $\pm$ 0.093 & 0.630 $\pm$ 0.002 $\pm$ 0.032 & 2.865 $\pm$ 0.017 $\pm$ 0.152 \\ \cline{1-5}
 2.775 & 1.620 $\pm$ 0.019 $\pm$ 0.042 $\pm$ 0.016 & 1.961 $\pm$ 0.015 $\pm$ 0.089 & 0.623 $\pm$ 0.008 $\pm$ 0.032 & 2.819 $\pm$ 0.062 $\pm$ 0.151 \\ \cline{1-5}
 2.825 &  			                   & 1.939 $\pm$ 0.029 $\pm$ 0.092 & 0.613 $\pm$ 0.007 $\pm$ 0.032 & 2.780 $\pm$ 0.038 $\pm$ 0.149 \\ \cline{1-5}
 2.875 &  			                   & 1.912 $\pm$ 0.015 $\pm$ 0.103 & 0.600 $\pm$ 0.008 $\pm$ 0.031 & 2.707 $\pm$ 0.060 $\pm$ 0.144 \\ \cline{1-5}
 2.925 &  			                   & 1.891 $\pm$ 0.005 $\pm$ 0.090 & 0.591 $\pm$ 0.003 $\pm$ 0.030 & 2.668 $\pm$ 0.020 $\pm$ 0.139 \\ \cline{1-5}
 2.975 &  			                   & 1.896 $\pm$ 0.016 $\pm$ 0.096 & 0.588 $\pm$ 0.007 $\pm$ 0.032 & 2.641 $\pm$ 0.057 $\pm$ 0.142 \\ \cline{1-5}
 3.025 & 1.624 $\pm$ 0.021 $\pm$ 0.045 $\pm$ 0.034 & 1.873 $\pm$ 0.011 $\pm$ 0.090 & 0.581 $\pm$ 0.006 $\pm$ 0.032 & 2.607 $\pm$ 0.057 $\pm$ 0.144 \\ \cline{1-5}
 3.075 &  			                   & 1.853 $\pm$ 0.039 $\pm$ 0.090 & 0.571 $\pm$ 0.007 $\pm$ 0.032 & 2.557 $\pm$ 0.038 $\pm$ 0.145 \\ \cline{1-5}
 3.125 &  			                   & 1.851 $\pm$ 0.006 $\pm$ 0.100 & 0.571 $\pm$ 0.003 $\pm$ 0.032 & 2.566 $\pm$ 0.020 $\pm$ 0.147 \\ \cline{1-5}
 3.175 &  			                   & 1.852 $\pm$ 0.031 $\pm$ 0.110 & 0.567 $\pm$ 0.009 $\pm$ 0.034 & 2.536 $\pm$ 0.059 $\pm$ 0.149 \\ \cline{1-5}
 3.225 &  			                   & 1.823 $\pm$ 0.007 $\pm$ 0.097 & 0.557 $\pm$ 0.003 $\pm$ 0.032 & 2.482 $\pm$ 0.021 $\pm$ 0.146 \\ \cline{1-5}
 3.275 &  			                   & 1.818 $\pm$ 0.018 $\pm$ 0.098 & 0.551 $\pm$ 0.009 $\pm$ 0.033 & 2.429 $\pm$ 0.067 $\pm$ 0.144 \\ \cline{1-5}
\end{tabular}
\end{ruledtabular}
\end{table*}
\begin{table*}
\begin{ruledtabular}
\begin{tabular}{|c|c|c|c|c|} \cline{1-5}
$Q^2~[$(GeV/c)$^2]$ & $M_2(Q^2)$x$10^{-1}$ & $M_4(Q^2)$x$10^{-2}$ & $M_6(Q^2)$x$10^{-2}$ & $M_8(Q^2)$x$10^{-3}$ \\ \cline{1-5}
 3.325 & 1.632 $\pm$ 0.029 $\pm$ 0.052 $\pm$ 0.036 & 1.816 $\pm$ 0.016 $\pm$ 0.095 & 0.545 $\pm$ 0.008 $\pm$ 0.032 & 2.397 $\pm$ 0.071 $\pm$ 0.144 \\ \cline{1-5}
 3.375 &  			                   & 1.807 $\pm$ 0.017 $\pm$ 0.090 & 0.544 $\pm$ 0.009 $\pm$ 0.032 & 2.390 $\pm$ 0.069 $\pm$ 0.144 \\ \cline{1-5}
 3.425 &  			                   & 1.779 $\pm$ 0.005 $\pm$ 0.101 & 0.536 $\pm$ 0.003 $\pm$ 0.031 & 2.367 $\pm$ 0.018 $\pm$ 0.144 \\ \cline{1-5}
 3.475 & 1.622 $\pm$ 0.007 $\pm$ 0.035 $\pm$ 0.025 & 1.792 $\pm$ 0.005 $\pm$ 0.090 & 0.531 $\pm$ 0.002 $\pm$ 0.032 & 2.314 $\pm$ 0.019 $\pm$ 0.143 \\ \cline{1-5}
 3.525 &  			                   & 1.754 $\pm$ 0.023 $\pm$ 0.090 & 0.527 $\pm$ 0.010 $\pm$ 0.031 & 2.298 $\pm$ 0.067 $\pm$ 0.143 \\ \cline{1-5}
 3.575 &  			                   & 1.742 $\pm$ 0.015 $\pm$ 0.083 & 0.518 $\pm$ 0.003 $\pm$ 0.029 & 2.272 $\pm$ 0.020 $\pm$ 0.138 \\ \cline{1-5}
 3.625 &  			                   & 1.758 $\pm$ 0.030 $\pm$ 0.120 & 0.514 $\pm$ 0.009 $\pm$ 0.033 & 2.224 $\pm$ 0.054 $\pm$ 0.138 \\ \cline{1-5}
 3.675 &  			                   & 1.717 $\pm$ 0.030 $\pm$ 0.093 & 0.509 $\pm$ 0.005 $\pm$ 0.032 & 2.210 $\pm$ 0.029 $\pm$ 0.138 \\ \cline{1-5}
 3.725 &  			                   & 1.739 $\pm$ 0.030 $\pm$ 0.130 & 0.508 $\pm$ 0.009 $\pm$ 0.034 & 2.196 $\pm$ 0.060 $\pm$ 0.146 \\ \cline{1-5}
 3.775 &  			                   & 1.703 $\pm$ 0.023 $\pm$ 0.094 & 0.504 $\pm$ 0.005 $\pm$ 0.033 & 2.188 $\pm$ 0.025 $\pm$ 0.148 \\ \cline{1-5}
 3.825 &  			                   & 1.735 $\pm$ 0.006 $\pm$ 0.126 & 0.501 $\pm$ 0.003 $\pm$ 0.034 & 2.165 $\pm$ 0.025 $\pm$ 0.142 \\ \cline{1-5}
 3.875 &  			                   & 1.681 $\pm$ 0.021 $\pm$ 0.105 & 0.498 $\pm$ 0.003 $\pm$ 0.031 & 2.179 $\pm$ 0.020 $\pm$ 0.140 \\ \cline{1-5}
 3.925 &  			                   &                               & 0.496 $\pm$ 0.009 $\pm$ 0.034 & 2.145 $\pm$ 0.051 $\pm$ 0.143 \\ \cline{1-5}
 3.975 &  			                   &                               & 0.494 $\pm$ 0.002 $\pm$ 0.030 & 2.151 $\pm$ 0.016 $\pm$ 0.139 \\ \cline{1-5}
 4.025 &  			                   & 1.694 $\pm$ 0.031 $\pm$ 0.085 & 0.494 $\pm$ 0.010 $\pm$ 0.031 & 2.123 $\pm$ 0.066 $\pm$ 0.137 \\ \cline{1-5}
 4.075 &  			                   &                               & 0.487 $\pm$ 0.006 $\pm$ 0.034 & 2.101 $\pm$ 0.030 $\pm$ 0.143 \\ \cline{1-5}
 4.125 &  			                   & 1.657 $\pm$ 0.009 $\pm$ 0.083 & 0.486 $\pm$ 0.003 $\pm$ 0.031 & 2.121 $\pm$ 0.017 $\pm$ 0.144 \\ \cline{1-5}
 4.175 &  			                   & 1.658 $\pm$ 0.062 $\pm$ 0.086 & 0.486 $\pm$ 0.013 $\pm$ 0.031 & 2.112 $\pm$ 0.073 $\pm$ 0.146 \\ \cline{1-5}
 4.225 &  			                   & 1.664 $\pm$ 0.045 $\pm$ 0.093 & 0.482 $\pm$ 0.014 $\pm$ 0.031 & 2.084 $\pm$ 0.097 $\pm$ 0.147 \\ \cline{1-5}
 4.275 &  			                   & 1.630 $\pm$ 0.030 $\pm$ 0.075 & 0.475 $\pm$ 0.010 $\pm$ 0.030 & 2.041 $\pm$ 0.066 $\pm$ 0.141 \\ \cline{1-5}
 4.325 &  			                   & 1.636 $\pm$ 0.012 $\pm$ 0.085 & 0.474 $\pm$ 0.003 $\pm$ 0.031 & 2.047 $\pm$ 0.020 $\pm$ 0.146 \\ \cline{1-5}
 4.375 &  			                   &                               & 0.466 $\pm$ 0.007 $\pm$ 0.034 & 1.986 $\pm$ 0.036 $\pm$ 0.144 \\ \cline{1-5}
 4.425 &  			                   &                               & 0.465 $\pm$ 0.008 $\pm$ 0.035 & 1.982 $\pm$ 0.050 $\pm$ 0.147 \\ \cline{1-5}
 4.475 &  			                   &                               & 0.459 $\pm$ 0.005 $\pm$ 0.035 & 1.938 $\pm$ 0.038 $\pm$ 0.149 \\ \cline{1-5}
 4.525 &  			                   & 1.612 $\pm$ 0.022 $\pm$ 0.076 & 0.457 $\pm$ 0.004 $\pm$ 0.029 & 1.941 $\pm$ 0.019 $\pm$ 0.143 \\ \cline{1-5}
 4.575 &  			                   & 1.602 $\pm$ 0.048 $\pm$ 0.076 & 0.456 $\pm$ 0.011 $\pm$ 0.029 & 1.940 $\pm$ 0.060 $\pm$ 0.144 \\ \cline{1-5}
 4.625 &  			                   &                               & 0.456 $\pm$ 0.011 $\pm$ 0.037 & 1.927 $\pm$ 0.058 $\pm$ 0.157 \\ \cline{1-5}
 4.675 &  			                   &                               & 0.447 $\pm$ 0.006 $\pm$ 0.030 & 1.898 $\pm$ 0.034 $\pm$ 0.145 \\ \cline{1-5}
 4.725 &  			                   &                               & 0.443 $\pm$ 0.004 $\pm$ 0.031 & 1.881 $\pm$ 0.019 $\pm$ 0.146 \\ \cline{1-5}
 4.775 &  			                   &                               & 0.441 $\pm$ 0.006 $\pm$ 0.035 & 1.839 $\pm$ 0.032 $\pm$ 0.150 \\ \cline{1-5}
 4.825 &  			                   &                               & 0.442 $\pm$ 0.007 $\pm$ 0.035 & 1.847 $\pm$ 0.037 $\pm$ 0.152 \\ \cline{1-5}
 4.875 &  			                   & 1.556 $\pm$ 0.050 $\pm$ 0.071 & 0.433 $\pm$ 0.006 $\pm$ 0.029 & 1.811 $\pm$ 0.021 $\pm$ 0.144 \\ \cline{1-5}
 4.925 &  			                   &                               &                               & 1.793 $\pm$ 0.016 $\pm$ 0.162 \\ \cline{1-5}
 4.975 &  			                   &                               & 0.430 $\pm$ 0.004 $\pm$ 0.027 & 1.795 $\pm$ 0.019 $\pm$ 0.138 \\ \cline{1-5}
 5.025 &  			                   &                               & 0.430 $\pm$ 0.012 $\pm$ 0.026 & 1.795 $\pm$ 0.063 $\pm$ 0.137 \\ \cline{1-5}
 5.075 &  			                   &                               &                               & 1.781 $\pm$ 0.018 $\pm$ 0.151 \\ \cline{1-5}
 5.125 &  			                   & 1.499 $\pm$ 0.017 $\pm$ 0.032 &                               & 1.756 $\pm$ 0.059 $\pm$ 0.146 \\ \cline{1-5}
 5.175 &  			                   &                               &                               & 1.753 $\pm$ 0.016 $\pm$ 0.125 \\ \cline{1-5}
 5.225 &  			                   &                               &                               & 1.774 $\pm$ 0.050 $\pm$ 0.154 \\ \cline{1-5}
 5.275 &  			                   &                               & 0.418 $\pm$ 0.010 $\pm$ 0.027 & 1.737 $\pm$ 0.038 $\pm$ 0.136 \\ \cline{1-5}
 5.325 &  			                   &                               & 0.419 $\pm$ 0.014 $\pm$ 0.024 & 1.739 $\pm$ 0.059 $\pm$ 0.133 \\ \cline{1-5}
 5.375 &  			                   &                               & 0.416 $\pm$ 0.005 $\pm$ 0.024 & 1.710 $\pm$ 0.021 $\pm$ 0.125 \\ \cline{1-5}
 5.425 &  			                   &                               &                               & 1.706 $\pm$ 0.030 $\pm$ 0.149 \\ \cline{1-5}
 5.475 & 1.545 $\pm$ 0.013 $\pm$ 0.032 $\pm$ 0.038 &                               &                               & 1.704 $\pm$ 0.054 $\pm$ 0.154 \\ \cline{1-5}
 5.525 &  			                   &                               &                               & 1.647 $\pm$ 0.033 $\pm$ 0.144 \\ \cline{1-5}
 5.625 &  			                   &                               & 0.404 $\pm$ 0.003 $\pm$ 0.020 & 1.643 $\pm$ 0.020 $\pm$ 0.113 \\ \cline{1-5}
 5.925 &  			                   &                               &                               & 1.603 $\pm$ 0.026 $\pm$ 0.098 \\ \cline{1-5}
 5.955 &  			                   & 1.436 $\pm$ 0.023 $\pm$ 0.027 & 0.374 $\pm$ 0.008 $\pm$ 0.011 & 1.472 $\pm$ 0.032 $\pm$ 0.052 \\ \cline{1-5}
 6.915 & 1.521 $\pm$ 0.009 $\pm$ 0.014 $\pm$ 0.033 & 1.404 $\pm$ 0.011 $\pm$ 0.027 & 0.361 $\pm$ 0.004 $\pm$ 0.010 & 1.389 $\pm$ 0.020 $\pm$ 0.044 \\ \cline{1-5}
 7.267 &  			                   & 1.376 $\pm$ 0.012 $\pm$ 0.036 & 0.353 $\pm$ 0.004 $\pm$ 0.012 & 1.363 $\pm$ 0.022 $\pm$ 0.065 \\ \cline{1-5}
 7.630 &  			                   &                               & 0.343 $\pm$ 0.004 $\pm$ 0.019 & 1.308 $\pm$ 0.022 $\pm$ 0.125 \\ \cline{1-5}
 8.021 &  			                   &                               & 0.336 $\pm$ 0.002 $\pm$ 0.012 & 1.274 $\pm$ 0.010 $\pm$ 0.052 \\ \cline{1-5}
 8.847 & 1.508 $\pm$ 0.013 $\pm$ 0.015 $\pm$ 0.041 & 1.325 $\pm$ 0.011 $\pm$ 0.027 & 0.329 $\pm$ 0.003 $\pm$ 0.010 & 1.215 $\pm$ 0.017 $\pm$ 0.044 \\ \cline{1-5}
 9.775 &  			                   & 1.281 $\pm$ 0.005 $\pm$ 0.038 & 0.313 $\pm$ 0.001 $\pm$ 0.010 & 1.146 $\pm$ 0.005 $\pm$ 0.041 \\ \cline{1-5}
10.267 &  			                   &                               &                               & 1.156 $\pm$ 0.010 $\pm$ 0.031 \\ \cline{1-5}
10.762 &  			                   &                               & 0.306 $\pm$ 0.002 $\pm$ 0.007 & 1.115 $\pm$ 0.011 $\pm$ 0.029 \\ \cline{1-5}
11.344 & 1.500 $\pm$ 0.014 $\pm$ 0.010 $\pm$ 0.046 & 1.242 $\pm$ 0.015 $\pm$ 0.021 & 0.299 $\pm$ 0.005 $\pm$ 0.008 & 1.084 $\pm$ 0.035 $\pm$ 0.033 \\ \cline{1-5}
12.580 &  			                   &                               &                               & 1.049 $\pm$ 0.013 $\pm$ 0.022 \\ \cline{1-5}
13.238 & 1.503 $\pm$ 0.025 $\pm$ 0.023 $\pm$ 0.049 & 1.205 $\pm$ 0.011 $\pm$ 0.019 & 0.287 $\pm$ 0.003 $\pm$ 0.006 & 1.001 $\pm$ 0.010 $\pm$ 0.026 \\ \cline{1-5}
14.689 & 1.520 $\pm$ 0.012 $\pm$ 0.011 $\pm$ 0.052 & 1.216 $\pm$ 0.013 $\pm$ 0.027 & 0.285 $\pm$ 0.003 $\pm$ 0.010 & 0.992 $\pm$ 0.013 $\pm$ 0.048 \\ \cline{1-5}
17.108 & 1.478 $\pm$ 0.011 $\pm$ 0.017 $\pm$ 0.052 & 1.112 $\pm$ 0.011 $\pm$ 0.038 & 0.256 $\pm$ 0.003 $\pm$ 0.011 & 0.900 $\pm$ 0.015 $\pm$ 0.039 \\ \cline{1-5}
\end{tabular}
\end{ruledtabular}
\end{table*}
\begin{table*}
\begin{ruledtabular}
\begin{tabular}{|c|c|c|c|c|} \cline{1-5}
$Q^2~[$(GeV/c)$^2]$ & $M_2(Q^2)$x$10^{-1}$ & $M_4(Q^2)$x$10^{-2}$ & $M_6(Q^2)$x$10^{-2}$ & $M_8(Q^2)$x$10^{-3}$ \\ \cline{1-5}
19.072 &  			                   & 1.107 $\pm$ 0.010 $\pm$ 0.021 & 0.257 $\pm$ 0.003 $\pm$ 0.007 & 0.911 $\pm$ 0.014 $\pm$ 0.028 \\ \cline{1-5}
20.108 & 1.467 $\pm$ 0.010 $\pm$ 0.007 $\pm$ 0.058 & 1.094 $\pm$ 0.014 $\pm$ 0.019 & 0.250 $\pm$ 0.004 $\pm$ 0.008 & 0.874 $\pm$ 0.016 $\pm$ 0.039 \\ \cline{1-5}
21.097 &*1.486 $\pm$ 0.010 $\pm$ 0.011 $\pm$ 0.049 & 1.101 $\pm$ 0.012 $\pm$ 0.032 &				  &				  \\ \cline{1-5}
24.259 &*1.470 $\pm$ 0.012 $\pm$ 0.014 $\pm$ 0.051 & 1.052 $\pm$ 0.010 $\pm$ 0.040 &				  &				  \\ \cline{1-5}
26.680 &*1.469 $\pm$ 0.013 $\pm$ 0.007 $\pm$ 0.059 & 1.056 $\pm$ 0.010 $\pm$ 0.021 &				  &				  \\ \cline{1-5}
32.500 &*1.479 $\pm$ 0.013 $\pm$ 0.014 $\pm$ 0.055 &				  &				  &				  \\ \cline{1-5}
34.932 &*1.445 $\pm$ 0.113 $\pm$ 0.007 $\pm$ 0.064 &				  &				  &				  \\ \cline{1-5}
36.750 &*1.482 $\pm$ 0.012 $\pm$ 0.018 $\pm$ 0.057 & 1.014 $\pm$ 0.010 $\pm$ 0.016 & 0.223 $\pm$ 0.003 $\pm$ 0.004 & 0.767 $\pm$ 0.014 $\pm$ 0.014 \\ \cline{1-5}
43.970 &*1.478 $\pm$ 0.026 $\pm$ 0.010 $\pm$ 0.057 &				  &				  &				  \\ \cline{1-5}
47.440 &*1.477 $\pm$ 0.127 $\pm$ 0.007 $\pm$ 0.051 & 1.016 $\pm$ 0.029 $\pm$ 0.018 &				  &				  \\ \cline{1-5}
64.270 &  			                   & 9.450 $\pm$ 0.042 $\pm$ 0.014 & 0.206 $\pm$ 0.007 $\pm$ 0.003 & 0.699 $\pm$ 0.022 $\pm$ 0.010 \\ \cline{1-5}
75.000 &  			                   &*9.443 $\pm$ 0.017 $\pm$ 0.024 &*0.206 $\pm$ 0.006 $\pm$ 0.008 &			       \\ \cline{1-5}
86.000 &  			                   &*9.183 $\pm$ 0.018 $\pm$ 0.019 &*0.196 $\pm$ 0.007 $\pm$ 0.007 &			       \\ \cline{1-5}
97.690 &  			                   &*9.232 $\pm$ 0.019 $\pm$ 0.010 &*0.199 $\pm$ 0.005 $\pm$ 0.003 &*0.672 $\pm$ 0.023 $\pm$ 0.008 \\ \cline{1-5}
\end{tabular}
\end{ruledtabular}
\end{table*}

\begin{table*}
\caption{\label{table:twist1}Extracted parameters of the twist
expansion at the reference scale $\mu^2=10$ (GeV/c)$^2$.
The first uncertainty is the systematic one described in text, while
the second uncertainty has a statistical origin and is obtained from a
MINOS~\cite{cernlib} minimization procedure.
The contribution of twists-6 to $M_2$ was
too small to be extracted by the present procedure.}
\begin{ruledtabular}
\begin{tabular}{|c|c|c|c|c|} \hline
               & $M_2$                            & $M_4$                                     & $M_6$                                   & $M_8$                                      \\ \hline
$\eta_n(\mu^2)$& 0.152$\pm$0.02$\pm$0.03          & (1.215$\pm$0.03$\pm$0.01)$\times 10^{-2}$ & (2.95$\pm$0.1$\pm$0.04)$\times 10^{-3}$ & (1.05$\pm$0.02$\pm$0.02)$\times 10^{-3}$    \\ \hline
$a^{(4)}$      & (4$\pm$2$\pm$22)$\times 10^{-4}$ & (7.4$\pm$2$\pm$2.5)$\times 10^{-3}$       & (2.7$\pm$0.5$\pm$0.04)$\times 10^{-3}$  & (1.7$\pm$0.8$\pm$0.04)$\times 10^{-3}$ \\ \hline
$\gamma^{(4)}$ & 3.4$\pm$0.2$\pm$5.2              &    3$\pm$0.5$\pm$1                        & 5.9$\pm$0.3$\pm$0.02                    &  6.4$\pm$3.5$\pm$0.04                  \\ \hline
$a^{(6)}$      & -                                &(-1.5$\pm$0.2$\pm$0.3)$\times 10^{-2}$     & (-9.2$\pm$1.4$\pm$0.15)$\times 10^{-3}$ & (-6.6$\pm$2$\pm$0.16)$\times 10^{-3}$  \\ \hline
$\gamma^{(6)}$ & -                                &  1.9$\pm$0.4$\pm$0.8                      & 4.3$\pm$0.3$\pm$0.02                    &   4.7$\pm$1.4$\pm$0.04                 \\ \hline
\end{tabular}
\end{ruledtabular}
\end{table*}

\begin{table*}
\caption{\label{table:ltw} The extracted leading twist contribution
$\eta_n(Q^2)$
(see Eq.~\ref{eq:SGR})
shown in Fig.~\ref{fig:twists}, reported
with systematic uncertainties.}
\begin{ruledtabular}
\begin{tabular}{|c|c|c|c|c|} \cline{1-5}
$Q^2~[$(GeV/c)$^2]$ & $\eta_2(Q^2)$x$10^{-1}$ & $\eta_4(Q^2)$x$10^{-2}$ & $\eta_6(Q^2)$x$10^{-3}$ & $\eta_8(Q^2)$x$10^{-3}$ \\ \cline{1-5}
  1.025 & 1.84 $\pm$ 0.07 & 2.56 $\pm$ 0.06 & 11.5 $\pm$ 0.4  & 8.63 $\pm$ 0.07 \\ \cline{1-5}
  1.075 & 1.83 $\pm$ 0.07 & 2.49 $\pm$ 0.05 & 10.6 $\pm$ 0.4  & 7.28 $\pm$ 0.06 \\ \cline{1-5}
  1.125 & 1.82 $\pm$ 0.07 & 2.42 $\pm$ 0.05 & 9.88 $\pm$ 0.3  & 6.34 $\pm$ 0.05 \\ \cline{1-5}
  1.175 & 1.81 $\pm$ 0.07 & 2.36 $\pm$ 0.05 & 9.29 $\pm$ 0.3  & 5.64 $\pm$ 0.05 \\ \cline{1-5}
  1.225 & 1.80 $\pm$ 0.07 & 2.31 $\pm$ 0.05 & 8.80 $\pm$ 0.3  & 5.11 $\pm$ 0.04 \\ \cline{1-5}
  1.275 & 1.79 $\pm$ 0.07 & 2.26 $\pm$ 0.05 & 8.38 $\pm$ 0.3  & 4.68 $\pm$ 0.04 \\ \cline{1-5}
  1.325 & 1.78 $\pm$ 0.06 & 2.21 $\pm$ 0.05 & 8.02 $\pm$ 0.3  & 4.34 $\pm$ 0.04 \\ \cline{1-5}
  1.375 & 1.77 $\pm$ 0.06 & 2.17 $\pm$ 0.05 & 7.70 $\pm$ 0.3  & 4.05 $\pm$ 0.03 \\ \cline{1-5}
  1.425 & 1.77 $\pm$ 0.06 & 2.13 $\pm$ 0.05 & 7.43 $\pm$ 0.3  & 3.82 $\pm$ 0.03 \\ \cline{1-5}
  1.475 & 1.76 $\pm$ 0.06 & 2.10 $\pm$ 0.05 & 7.18 $\pm$ 0.3  & 3.61 $\pm$ 0.03 \\ \cline{1-5}
  1.525 & 1.75 $\pm$ 0.06 & 2.07 $\pm$ 0.05 & 6.96 $\pm$ 0.2  & 3.43 $\pm$ 0.03 \\ \cline{1-5}
  1.575 & 1.74 $\pm$ 0.06 & 2.04 $\pm$ 0.04 & 6.76 $\pm$ 0.2  & 3.28 $\pm$ 0.03 \\ \cline{1-5}
  1.625 & 1.74 $\pm$ 0.06 & 2.01 $\pm$ 0.04 & 6.58 $\pm$ 0.2  & 3.14 $\pm$ 0.03 \\ \cline{1-5}
  1.675 & 1.73 $\pm$ 0.06 & 1.98 $\pm$ 0.04 & 6.41 $\pm$ 0.2  & 3.02 $\pm$ 0.02 \\ \cline{1-5}
  1.725 & 1.72 $\pm$ 0.06 & 1.95 $\pm$ 0.04 & 6.26 $\pm$ 0.2  & 2.91 $\pm$ 0.02 \\ \cline{1-5}
  1.775 & 1.72 $\pm$ 0.06 & 1.93 $\pm$ 0.04 & 6.12 $\pm$ 0.2  & 2.81 $\pm$ 0.02 \\ \cline{1-5}
  1.825 & 1.71 $\pm$ 0.05 & 1.91 $\pm$ 0.04 & 6.00 $\pm$ 0.2  & 2.73 $\pm$ 0.02 \\ \cline{1-5}
  1.875 & 1.70 $\pm$ 0.05 & 1.89 $\pm$ 0.04 & 5.88 $\pm$ 0.2  & 2.65 $\pm$ 0.02 \\ \cline{1-5}
  1.925 & 1.70 $\pm$ 0.05 & 1.87 $\pm$ 0.04 & 5.76 $\pm$ 0.2  & 2.57 $\pm$ 0.02 \\ \cline{1-5}
  1.975 & 1.69 $\pm$ 0.05 & 1.85 $\pm$ 0.04 & 5.66 $\pm$ 0.2  & 2.50 $\pm$ 0.02 \\ \cline{1-5}
  2.025 & 1.69 $\pm$ 0.05 & 1.83 $\pm$ 0.04 & 5.56 $\pm$ 0.2  & 2.44 $\pm$ 0.02 \\ \cline{1-5}
  2.075 & 1.68 $\pm$ 0.05 & 1.81 $\pm$ 0.04 & 5.47 $\pm$ 0.2  & 2.38 $\pm$ 0.02 \\ \cline{1-5}
  2.125 & 1.68 $\pm$ 0.05 & 1.80 $\pm$ 0.04 & 5.39 $\pm$ 0.2  & 2.33 $\pm$ 0.02 \\ \cline{1-5}
  2.175 & 1.67 $\pm$ 0.05 & 1.78 $\pm$ 0.04 & 5.31 $\pm$ 0.2  & 2.28 $\pm$ 0.02 \\ \cline{1-5}
  2.225 & 1.67 $\pm$ 0.05 & 1.77 $\pm$ 0.04 & 5.23 $\pm$ 0.2  & 2.23 $\pm$ 0.02 \\ \cline{1-5}
  2.275 & 1.66 $\pm$ 0.05 & 1.75 $\pm$ 0.04 & 5.16 $\pm$ 0.2  & 2.19 $\pm$ 0.02 \\ \cline{1-5}
  2.325 & 1.66 $\pm$ 0.05 & 1.74 $\pm$ 0.04 & 5.11 $\pm$ 0.2  & 2.16 $\pm$ 0.02 \\ \cline{1-5}
  2.375 & 1.66 $\pm$ 0.05 & 1.73 $\pm$ 0.04 & 5.05 $\pm$ 0.2  & 2.12 $\pm$ 0.02 \\ \cline{1-5}
  2.425 & 1.66 $\pm$ 0.05 & 1.72 $\pm$ 0.04 & 5.00 $\pm$ 0.2  & 2.09 $\pm$ 0.02 \\ \cline{1-5}
  2.475 & 1.65 $\pm$ 0.05 & 1.71 $\pm$ 0.04 & 4.94 $\pm$ 0.2  & 2.06 $\pm$ 0.02 \\ \cline{1-5}
  2.525 & 1.65 $\pm$ 0.05 & 1.70 $\pm$ 0.04 & 4.89 $\pm$ 0.2  & 2.04 $\pm$ 0.02 \\ \cline{1-5}
  2.575 & 1.65 $\pm$ 0.05 & 1.69 $\pm$ 0.04 & 4.85 $\pm$ 0.2  & 2.01 $\pm$ 0.02 \\ \cline{1-5}
  2.625 & 1.64 $\pm$ 0.05 & 1.68 $\pm$ 0.04 & 4.80 $\pm$ 0.2  & 1.98 $\pm$ 0.02 \\ \cline{1-5}
  2.675 & 1.64 $\pm$ 0.04 & 1.67 $\pm$ 0.04 & 4.76 $\pm$ 0.2  & 1.96 $\pm$ 0.02 \\ \cline{1-5}
  2.725 & 1.64 $\pm$ 0.04 & 1.66 $\pm$ 0.04 & 4.71 $\pm$ 0.2  & 1.93 $\pm$ 0.02 \\ \cline{1-5}
  2.775 & 1.64 $\pm$ 0.04 & 1.65 $\pm$ 0.04 & 4.67 $\pm$ 0.2  & 1.91 $\pm$ 0.02 \\ \cline{1-5}
  2.825 & 1.64 $\pm$ 0.04 & 1.64 $\pm$ 0.04 & 4.64 $\pm$ 0.2  & 1.89 $\pm$ 0.02 \\ \cline{1-5}
  2.875 & 1.63 $\pm$ 0.04 & 1.63 $\pm$ 0.04 & 4.60 $\pm$ 0.2  & 1.87 $\pm$ 0.02 \\ \cline{1-5}
  2.925 & 1.63 $\pm$ 0.04 & 1.62 $\pm$ 0.04 & 4.56 $\pm$ 0.2  & 1.85 $\pm$ 0.02 \\ \cline{1-5}
  2.975 & 1.63 $\pm$ 0.04 & 1.62 $\pm$ 0.04 & 4.53 $\pm$ 0.2  & 1.83 $\pm$ 0.01 \\ \cline{1-5}
  3.025 & 1.63 $\pm$ 0.04 & 1.61 $\pm$ 0.04 & 4.49 $\pm$ 0.2  & 1.81 $\pm$ 0.01 \\ \cline{1-5}
  3.075 & 1.63 $\pm$ 0.04 & 1.60 $\pm$ 0.04 & 4.46 $\pm$ 0.2  & 1.79 $\pm$ 0.01 \\ \cline{1-5}
  3.125 & 1.62 $\pm$ 0.04 & 1.59 $\pm$ 0.04 & 4.43 $\pm$ 0.2  & 1.78 $\pm$ 0.01 \\ \cline{1-5}
  3.175 & 1.62 $\pm$ 0.04 & 1.59 $\pm$ 0.03 & 4.40 $\pm$ 0.2  & 1.76 $\pm$ 0.01 \\ \cline{1-5}
  3.225 & 1.62 $\pm$ 0.04 & 1.58 $\pm$ 0.03 & 4.36 $\pm$ 0.2  & 1.74 $\pm$ 0.01 \\ \cline{1-5}
  3.275 & 1.62 $\pm$ 0.04 & 1.57 $\pm$ 0.03 & 4.34 $\pm$ 0.2  & 1.73 $\pm$ 0.01 \\ \cline{1-5}
  3.325 & 1.62 $\pm$ 0.04 & 1.57 $\pm$ 0.03 & 4.31 $\pm$ 0.2  & 1.71 $\pm$ 0.01 \\ \cline{1-5}
  3.375 & 1.61 $\pm$ 0.04 & 1.56 $\pm$ 0.03 & 4.28 $\pm$ 0.2  & 1.70 $\pm$ 0.01 \\ \cline{1-5}
  3.425 & 1.61 $\pm$ 0.04 & 1.55 $\pm$ 0.03 & 4.25 $\pm$ 0.1  & 1.68 $\pm$ 0.01 \\ \cline{1-5}
  3.475 & 1.61 $\pm$ 0.04 & 1.55 $\pm$ 0.03 & 4.23 $\pm$ 0.1  & 1.67 $\pm$ 0.01 \\ \cline{1-5}
  3.525 & 1.61 $\pm$ 0.04 & 1.54 $\pm$ 0.03 & 4.20 $\pm$ 0.1  & 1.66 $\pm$ 0.01 \\ \cline{1-5}
  3.575 & 1.61 $\pm$ 0.04 & 1.53 $\pm$ 0.03 & 4.18 $\pm$ 0.1  & 1.64 $\pm$ 0.01 \\ \cline{1-5}
  3.625 & 1.61 $\pm$ 0.04 & 1.53 $\pm$ 0.03 & 4.15 $\pm$ 0.1  & 1.63 $\pm$ 0.01 \\ \cline{1-5}
  3.675 & 1.61 $\pm$ 0.04 & 1.52 $\pm$ 0.03 & 4.13 $\pm$ 0.1  & 1.62 $\pm$ 0.01 \\ \cline{1-5}
  3.725 & 1.60 $\pm$ 0.04 & 1.52 $\pm$ 0.03 & 4.11 $\pm$ 0.1  & 1.61 $\pm$ 0.01 \\ \cline{1-5}
  3.775 & 1.60 $\pm$ 0.04 & 1.51 $\pm$ 0.03 & 4.08 $\pm$ 0.1  & 1.60 $\pm$ 0.01 \\ \cline{1-5}
  3.825 & 1.60 $\pm$ 0.04 & 1.51 $\pm$ 0.03 & 4.06 $\pm$ 0.1  & 1.59 $\pm$ 0.01 \\ \cline{1-5}
  3.875 & 1.60 $\pm$ 0.04 & 1.50 $\pm$ 0.03 & 4.04 $\pm$ 0.1  & 1.58 $\pm$ 0.01 \\ \cline{1-5}
  3.925 & 1.60 $\pm$ 0.04 & 1.50 $\pm$ 0.03 & 4.02 $\pm$ 0.1  & 1.56 $\pm$ 0.01 \\ \cline{1-5}
  3.975 & 1.60 $\pm$ 0.04 & 1.49 $\pm$ 0.03 & 4.00 $\pm$ 0.1  & 1.55 $\pm$ 0.01 \\ \cline{1-5}
\end{tabular}
\end{ruledtabular}
\end{table*}
\begin{table*}
\begin{ruledtabular}
\begin{tabular}{|c|c|c|c|c|} \cline{1-5}
$Q^2~[$(GeV/c)$^2]$ & $\eta_2(Q^2)$x$10^{-1}$ & $\eta_4(Q^2)$x$10^{-2}$ & $\eta_6(Q^2)$x$10^{-3}$ & $\eta_8(Q^2)$x$10^{-3}$ \\ \cline{1-5}
  4.025 & 1.60 $\pm$ 0.03 & 1.49 $\pm$ 0.03 & 3.98 $\pm$ 0.1  & 1.54 $\pm$ 0.01 \\ \cline{1-5}
  4.075 & 1.59 $\pm$ 0.03 & 1.48 $\pm$ 0.03 & 3.96 $\pm$ 0.1  & 1.53 $\pm$ 0.01 \\ \cline{1-5}
  4.125 & 1.59 $\pm$ 0.03 & 1.48 $\pm$ 0.03 & 3.94 $\pm$ 0.1  & 1.52 $\pm$ 0.01 \\ \cline{1-5}
  4.175 & 1.59 $\pm$ 0.03 & 1.47 $\pm$ 0.03 & 3.92 $\pm$ 0.1  & 1.52 $\pm$ 0.01 \\ \cline{1-5}
  4.225 & 1.59 $\pm$ 0.03 & 1.47 $\pm$ 0.03 & 3.91 $\pm$ 0.1  & 1.51 $\pm$ 0.01 \\ \cline{1-5}
  4.275 & 1.59 $\pm$ 0.03 & 1.46 $\pm$ 0.03 & 3.89 $\pm$ 0.1  & 1.50 $\pm$ 0.01 \\ \cline{1-5}
  4.325 & 1.59 $\pm$ 0.03 & 1.46 $\pm$ 0.03 & 3.87 $\pm$ 0.1  & 1.49 $\pm$ 0.01 \\ \cline{1-5}
  4.375 & 1.59 $\pm$ 0.03 & 1.46 $\pm$ 0.03 & 3.86 $\pm$ 0.1  & 1.48 $\pm$ 0.01 \\ \cline{1-5}
  4.425 & 1.59 $\pm$ 0.03 & 1.45 $\pm$ 0.03 & 3.84 $\pm$ 0.1  & 1.47 $\pm$ 0.01 \\ \cline{1-5}
  4.475 & 1.59 $\pm$ 0.03 & 1.45 $\pm$ 0.03 & 3.82 $\pm$ 0.1  & 1.46 $\pm$ 0.01 \\ \cline{1-5}
  4.525 & 1.58 $\pm$ 0.03 & 1.44 $\pm$ 0.03 & 3.81 $\pm$ 0.1  & 1.46 $\pm$ 0.01 \\ \cline{1-5}
  4.575 & 1.58 $\pm$ 0.03 & 1.44 $\pm$ 0.03 & 3.79 $\pm$ 0.1  & 1.45 $\pm$ 0.01 \\ \cline{1-5}
  4.625 & 1.58 $\pm$ 0.03 & 1.44 $\pm$ 0.03 & 3.78 $\pm$ 0.1  & 1.44 $\pm$ 0.01 \\ \cline{1-5}
  4.675 & 1.58 $\pm$ 0.03 & 1.43 $\pm$ 0.03 & 3.76 $\pm$ 0.1  & 1.43 $\pm$ 0.01 \\ \cline{1-5}
  4.725 & 1.58 $\pm$ 0.03 & 1.43 $\pm$ 0.03 & 3.75 $\pm$ 0.1  & 1.43 $\pm$ 0.01 \\ \cline{1-5}
  4.775 & 1.58 $\pm$ 0.03 & 1.43 $\pm$ 0.03 & 3.73 $\pm$ 0.1  & 1.42 $\pm$ 0.01 \\ \cline{1-5}
  4.825 & 1.58 $\pm$ 0.03 & 1.42 $\pm$ 0.03 & 3.72 $\pm$ 0.1  & 1.41 $\pm$ 0.01 \\ \cline{1-5}
  4.875 & 1.58 $\pm$ 0.03 & 1.42 $\pm$ 0.03 & 3.70 $\pm$ 0.1  & 1.40 $\pm$ 0.01 \\ \cline{1-5}
  4.925 & 1.58 $\pm$ 0.03 & 1.41 $\pm$ 0.03 & 3.69 $\pm$ 0.1  & 1.40 $\pm$ 0.01 \\ \cline{1-5}
  4.975 & 1.58 $\pm$ 0.03 & 1.41 $\pm$ 0.03 & 3.68 $\pm$ 0.1  & 1.39 $\pm$ 0.01 \\ \cline{1-5}
  5.025 & 1.57 $\pm$ 0.03 & 1.41 $\pm$ 0.03 & 3.66 $\pm$ 0.1  & 1.38 $\pm$ 0.01 \\ \cline{1-5}
  5.075 & 1.57 $\pm$ 0.03 & 1.40 $\pm$ 0.03 & 3.65 $\pm$ 0.1  & 1.38 $\pm$ 0.01 \\ \cline{1-5}
  5.125 & 1.57 $\pm$ 0.03 & 1.40 $\pm$ 0.03 & 3.64 $\pm$ 0.1  & 1.37 $\pm$ 0.01 \\ \cline{1-5}
  5.275 & 1.57 $\pm$ 0.03 & 1.39 $\pm$ 0.03 & 3.60 $\pm$ 0.1  & 1.35 $\pm$ 0.01 \\ \cline{1-5}
  5.325 & 1.57 $\pm$ 0.03 & 1.39 $\pm$ 0.03 & 3.59 $\pm$ 0.1  & 1.35 $\pm$ 0.01 \\ \cline{1-5}
  5.375 & 1.57 $\pm$ 0.03 & 1.39 $\pm$ 0.03 & 3.58 $\pm$ 0.1  & 1.34 $\pm$ 0.01 \\ \cline{1-5}
  5.475 & 1.57 $\pm$ 0.03 & 1.38 $\pm$ 0.03 & 3.55 $\pm$ 0.1  & 1.33 $\pm$ 0.01 \\ \cline{1-5}
  5.525 & 1.57 $\pm$ 0.03 & 1.38 $\pm$ 0.03 & 3.54 $\pm$ 0.1  & 1.33 $\pm$ 0.01 \\ \cline{1-5}
  5.625 & 1.56 $\pm$ 0.03 & 1.37 $\pm$ 0.03 & 3.52 $\pm$ 0.1  & 1.32 $\pm$ 0.01 \\ \cline{1-5}
  5.675 & 1.56 $\pm$ 0.03 & 1.37 $\pm$ 0.03 & 3.51 $\pm$ 0.1  & 1.31 $\pm$ 0.01 \\ \cline{1-5}
  5.725 & 1.56 $\pm$ 0.03 & 1.37 $\pm$ 0.03 & 3.50 $\pm$ 0.1  & 1.31 $\pm$ 0.01 \\ \cline{1-5}
  5.955 & 1.56 $\pm$ 0.03 & 1.35 $\pm$ 0.03 & 3.45 $\pm$ 0.1  & 1.28 $\pm$ 0.01 \\ \cline{1-5}
  6.915 & 1.55 $\pm$ 0.03 & 1.31 $\pm$ 0.03 & 3.29 $\pm$ 0.1  & 1.21 $\pm$ 0.01 \\ \cline{1-5}
  7.267 & 1.54 $\pm$ 0.02 & 1.30 $\pm$ 0.03 & 3.24 $\pm$ 0.1  & 1.18 $\pm$ 0.01 \\ \cline{1-5}
  7.630 & 1.54 $\pm$ 0.02 & 1.28 $\pm$ 0.03 & 3.19 $\pm$ 0.1  & 1.16 $\pm$ 0.01 \\ \cline{1-5}
  8.021 & 1.54 $\pm$ 0.02 & 1.27 $\pm$ 0.03 & 3.14 $\pm$ 0.1  & 1.14 $\pm$ 0.01 \\ \cline{1-5}
  8.847 & 1.53 $\pm$ 0.02 & 1.24 $\pm$ 0.03 & 3.05 $\pm$ 0.1  & 1.10 $\pm$ 0.01 \\ \cline{1-5}
  9.775 & 1.52 $\pm$ 0.02 & 1.22 $\pm$ 0.03 & 2.97 $\pm$ 0.1  & 1.06 $\pm$ 0.01 \\ \cline{1-5}
 10.267 & 1.52 $\pm$ 0.02 & 1.21 $\pm$ 0.03 & 2.92 $\pm$ 0.1  & 1.04 $\pm$ 0.01 \\ \cline{1-5}
 10.762 & 1.52 $\pm$ 0.02 & 1.20 $\pm$ 0.03 & 2.89 $\pm$ 0.1  & 1.02 $\pm$ 0.01 \\ \cline{1-5}
 11.344 & 1.51 $\pm$ 0.02 & 1.19 $\pm$ 0.03 & 2.85 $\pm$ 0.1  & 1.00 $\pm$ 0.01 \\ \cline{1-5}
 12.580 & 1.51 $\pm$ 0.02 & 1.16 $\pm$ 0.03 & 2.77 $\pm$ 0.1  & 0.97 $\pm$ 0.01 \\ \cline{1-5}
 13.238 & 1.50 $\pm$ 0.02 & 1.15 $\pm$ 0.03 & 2.73 $\pm$ 0.1  & 0.95 $\pm$ 0.01 \\ \cline{1-5}
 14.689 & 1.50 $\pm$ 0.02 & 1.13 $\pm$ 0.02 & 2.66 $\pm$ 0.09 & 0.92 $\pm$ 0.01 \\ \cline{1-5}
 17.108 & 1.49 $\pm$ 0.01 & 1.10 $\pm$ 0.02 & 2.56 $\pm$ 0.09 & 0.88 $\pm$ 0.01 \\ \cline{1-5}
 19.072 & 1.48 $\pm$ 0.01 & 1.08 $\pm$ 0.02 & 2.50 $\pm$ 0.09 & 0.85 $\pm$ 0.01 \\ \cline{1-5}
 20.108 & 1.48 $\pm$ 0.01 & 1.07 $\pm$ 0.02 & 2.47 $\pm$ 0.09 & 0.84 $\pm$ 0.01 \\ \cline{1-5}
 21.097 & 1.48 $\pm$ 0.01 & 1.06 $\pm$ 0.02 & 2.44 $\pm$ 0.09 & 0.83 $\pm$ 0.01 \\ \cline{1-5}
 24.259 & 1.47 $\pm$ 0.01 & 1.04 $\pm$ 0.02 & 2.36 $\pm$ 0.08 & 0.80 $\pm$ 0.01 \\ \cline{1-5}
 26.680 & 1.47 $\pm$ 0.01 & 1.03 $\pm$ 0.02 & 2.32 $\pm$ 0.08 & 0.78 $\pm$ 0.01 \\ \cline{1-5}
 32.500 & 1.46 $\pm$ 0.01 & 1.00 $\pm$ 0.02 & 2.23 $\pm$ 0.08 & 0.74 $\pm$ 0.01 \\ \cline{1-5}
 34.932 & 1.46 $\pm$ 0.01 & 0.99 $\pm$ 0.02 & 2.19 $\pm$ 0.08 & 0.73 $\pm$ 0.01 \\ \cline{1-5}
 36.750 & 1.46 $\pm$ 0.01 & 0.98 $\pm$ 0.02 & 2.17 $\pm$ 0.08 & 0.72 $\pm$ 0.01 \\ \cline{1-5}
 43.970 & 1.46 $\pm$ 0.01 & 0.96 $\pm$ 0.02 & 2.10 $\pm$ 0.07 & 0.69 $\pm$ 0.01 \\ \cline{1-5}
 47.440 & 1.46 $\pm$ 0.01 & 0.95 $\pm$ 0.02 & 2.07 $\pm$ 0.07 & 0.68 $\pm$ 0.01 \\ \cline{1-5}
 64.270 & 1.46 $\pm$ 0.01 & 0.91 $\pm$ 0.02 & 1.96 $\pm$ 0.07 & 0.63 $\pm$ 0.01 \\ \cline{1-5}
 75.000 & 1.45 $\pm$ 0.01 & 0.89 $\pm$ 0.02 & 1.91 $\pm$ 0.07 & 0.61 $\pm$ 0.01 \\ \cline{1-5}
 86.000 & 1.45 $\pm$ 0.01 & 0.88 $\pm$ 0.02 & 1.86 $\pm$ 0.07 & 0.60 $\pm$ 0.01 \\ \cline{1-5}
 97.690 & 1.45 $\pm$ 0.01 & 0.86 $\pm$ 0.02 & 1.83 $\pm$ 0.06 & 0.58 $\pm$ 0.01 \\ \cline{1-5}
\end{tabular}
\end{ruledtabular}
\end{table*}

\begin{figure}
\includegraphics[bb=1cm 4cm 20cm 23cm, scale=0.4]{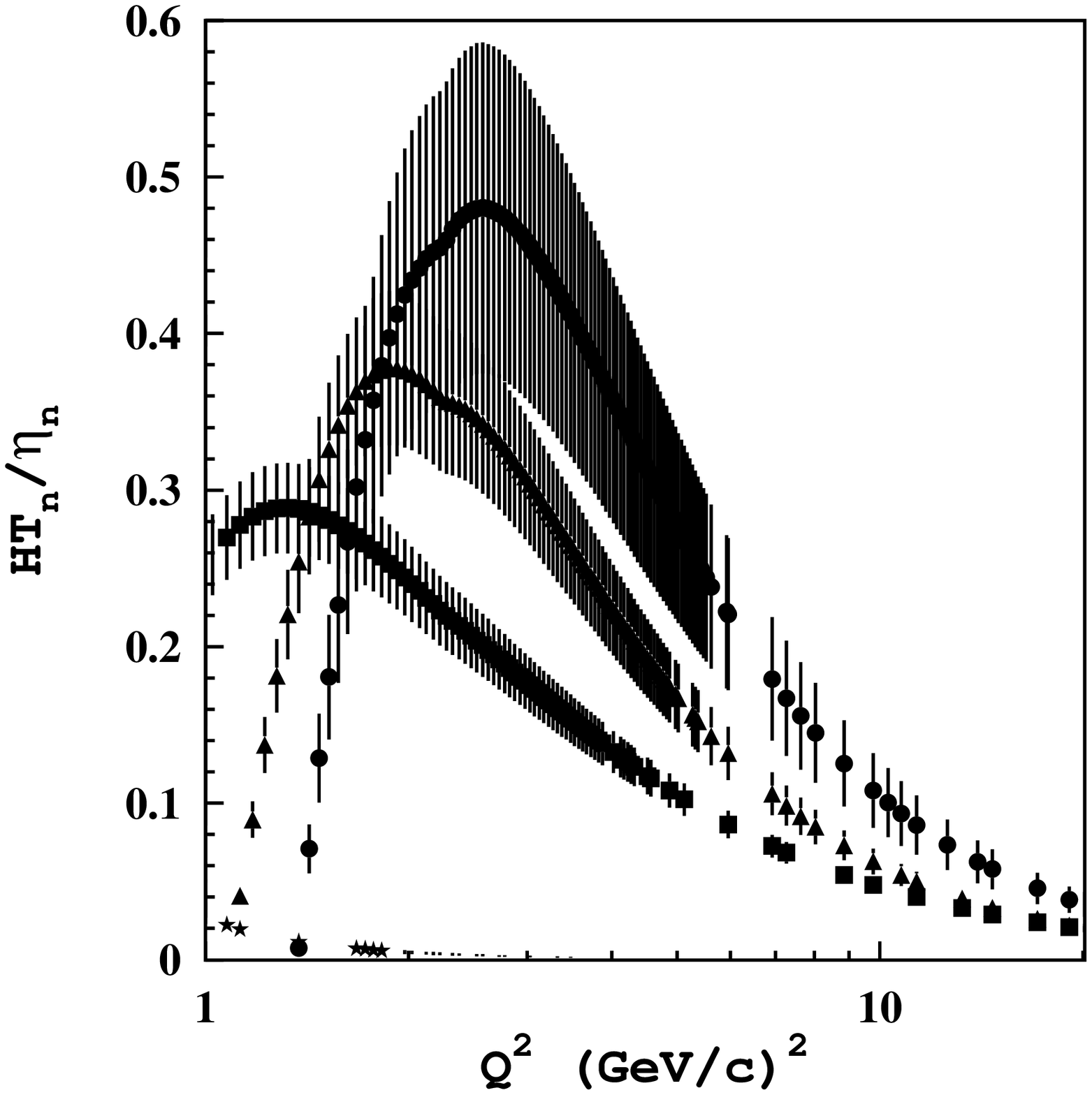}
\caption{\label{fig:ratio}Ratio of the total higher twist (see Eq.~\ref{eq:HT}) to
the leading twist given in Eq.~\ref{eq:SGR} with its systematic uncertainties.
Stars - $M_2$; squares - $M_4$;
triangles - $M_6$; circles - $M_8$.}
\end{figure}

Our results for each twist term are reported in Fig.~\ref{fig:twists} for
$n \geq 2$,
while the ratio of the total higher twist contribution to the
leading twist is shown in Fig.~\ref{fig:ratio}.
In addition, the extracted leading twist contribution is reported in
Table~\ref{table:ltw}.

\section{Conclusions}\label{sec:Conclusions}
We extracted the deuteron $F_2$ structure function in a continuous
two-dimensional range of $Q^2$ and $x$ from the inclusive
cross sections measured with CLAS.
The extracted structure functions are in good agreement
with previous measurements in overlapping regions, contribute
many additional kinematic points and sometimes improve the precision
where the world data exist.
Using these data, together with the previously available world data set,
we evaluated for the first time the Nachtmann moments $M_2(Q^2,x)$,
$M_4(Q^2,x)$,
$M_6(Q^2,x)$ and $M_8(Q^2,x)$ in the $Q^2$ range $0.5 - 100$~(GeV/c)$^2$.
Previously, the experimental information on the deuteron structure function
moments was missing in the low to medium $Q^2$ domain due to scarce data at
$x\rightarrow 1$.
Moreover, fixed $Q^2$ bins of the data render
the $Q^2$-evolution of the extracted moments model-independent.
The measured moments have been analyzed
in terms of a twist expansion in order to extract
both the leading and the higher twists simultaneously.
By calculating the $Q^2$-evolution of the leading twist at NLO and including
$\alpha_S$-higher order corrections through the soft gluon re-summation,
we extracted values of the reduced matrix elements $O_{n2}$
from Eq.~\ref{eq:i_m1}. Higher twists have been treated
phenomenologically within the pQCD-inspired
approach of Eq.~\ref{eq:HT} by introducing {\em effective} anomalous
dimensions. The $Q^2$ interval of the analysis was quite large, ranging
from 1 to 100~(GeV/c)$^2$, allowing us to determine the total
contribution of higher twists to the best accuracy possible.
The variation of the total higher twist contribution due to inclusion
of twist-8 and twist-10 terms is lower than the quoted systematic uncertainties.
The leading twist is determined with a few percent uncertainty, while the
precision of the higher twists decreases with $n$ reaching an overall 20-30\%
for $n=$8, thanks to the remarkable quality of the experimental
moments.

The main results of our twist analysis can be summarized as follows:
\begin{itemize}
\item the extracted leading term yields the dominant contribution
in the entire $Q^2$-range of the present analysis for all four moments.
This leads to the conclusion that despite the nuclear effects, a pQCD-based
description of the deuteron structure, including the effects of soft-gluon
re-summation, is surprisingly applicable also at low $Q^2$.
The corrections to the leading twist are significant but not crucial;
\item the $Q^2$-behaviour of the data indicates the presence of the
higher twist
contribution at $Q^2 <$5~(GeV/c)$^2$, positive at large $Q^2$ and negative at
$Q^2 \sim $1-2~(GeV/c)$^2$; the change of sign requires
in Eq.~\ref{eq:HT} at least two higher twist terms with opposite signs.
As already noted in Refs.~\cite{Ricco1,SIM00,osipenko_f2p},
such a cancellation makes the total higher twist contribution
smaller than its individual terms, which exceed the leading twist.
This partial {\em cancellation is a manifestation of the duality phenomena
in the pQCD representation}~\cite{BloGil}. It leads
to the prevailing DIS-inspired picture of virtual
photon-nucleon collisions also at low $Q^2$. The same mutual cancellation
of higher twist terms was observed in the proton structure function moments
in Refs.~\cite{osipenko_f2p,osipenko_g1p};
\item the total higher twist contribution is significant at
$Q^2 \approx$ few (GeV/c)$^2$ and large $x$.
This can be seen by comparing the higher twist contribution to
$M_8$, which is more heavily weighted in $x$, to $M_2$.
For $Q^2 > $6~(GeV/c)$^2$ the higher twist contribution does not
exceed $\simeq $20\% of the leading twist for all four moments.
\end{itemize}

Therefore, we have demonstrated that despite nuclear effects in the deuteron,
a pQCD-based analysis of the deuteron structure function moments is sensible,
so that a precise determination of the leading and higher twists is possible
with the new CLAS data.
The extracted values of the reduced matrix elements $O_{n2}$ still contain
some contribution of the nuclear off-shell and Fermi motion effects,
which should be taken care of before a comparison to Lattice QCD simulations
of the nucleon is made.
However, most of the nuclear effects, in particular FSI,
should be absorbed in the effective higher twist terms due to their scale
difference.
An estimate of the leading twist nuclear corrections would
allow extraction of the non-singlet part of the nucleon structure function
moments, which can be directly tested in the Lattice QCD simulations.

\begin{acknowledgments}
This work was supported by the Istituto Nazionale di Fisica Nucleare,
the French Commissariat \`a l'Energie Atomique, 
the French Centre National de la Recherche Scientifique,
the U.S. Department of Energy and National Science Foundation and
the Korea Science and Engineering Foundation.
The Southeastern Universities Research
Association (SURA) operates the Thomas Jefferson National Accelerator
Facility for the United States
Department of Energy under contract DE-AC05-84ER40150.
  
\end{acknowledgments}

\appendix

\section{Model of inclusive electron scattering cross section
off the deuteron and the deuteron structure function $F_2$}\label{app:f2_model}
\begin{figure}[!h]
\centering
\includegraphics[bb=2cm 4cm 18cm 24cm, scale=0.4]{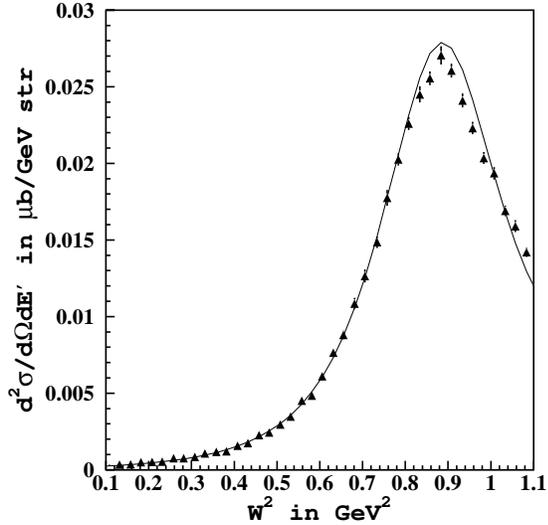}
\caption{Quasi-elastic cross section calculations in $\mu$b/(GeV~str) from
Ref.~\cite{Simula_QE}
compared to the experimental data from Ref.~\cite{E133} at $E=9.766$ GeV
and $\theta=10^\circ$. The uncertainties are statistical only.}
\label{fig:b_ics1}
\end{figure}

In order to extract efficiency, calculate radiative corrections
and evaluate moments of the structure function $F_2$,
it is essential to have a realistic model of the reaction
cross section. Thanks to many previous experiments, comprehensive knowledge
about electron-deuteron inclusive scattering is available. We based
our model on these previous results.
The model consists of three main elements:
\begin{itemize}
\item the elastic peak cross section was calculated using the deuteron elastic
form-factors from Ref.~\cite{Stuart_ed,Stuart_ed_ffs};
\item the quasi-elastic peak cross section was obtained within a model of
the nuclear structure of the deuteron~\cite{Simula_QE} using elementary
form-factors of
the proton and neutron from Ref.~\cite{Bosted}. This model is
based on the De Forest~\cite{deforest_cc1} cc1 prescription for the off-shell
nucleon and
includes various sophisticated treatments of the final state interactions.
Specifically designed for calculations of the quasi-elastic cross section
for light and complex nuclei, it reproduces existing data very
well (see Fig.~\ref{fig:b_ics1});
\item the inelastic cross section in the CLAS domain ($W^2<4.3$ GeV$^2$)
was taken from the fit to the recent Hall C data~\cite{f2-hc};
\item in DIS we have chosen the parameterization from Ref.~\cite{NMC97_f2_fit},
which describes particularly well the low-$x$ behaviour of the
structure function.
\end{itemize}
The elastic peak is very small in our $Q^2$ range and hence it is only
relevant for
the radiative correction calculations. The inelastic cross section model,
which fits the data from
Hall~C~\cite{f2-hc} very well, was obtained in the same kinematic domain.
The quasi-elastic cross section
calculations are model dependent and we have checked these before applying
them to the data.
Unfortunately, the Hall~C data do not contain the quasi-elastic peak, therefore
we had to compare to the previous SLAC and DESY measurements from
Refs.~\cite{E133,NE11,DESY}.
Some of these data are not corrected for the radiative corrections,
so we included radiative corrections in the model calculations.
An example of the comparison of the quasi-elastic cross section
model to the data, shown in Fig.~\ref{fig:b_ics1}, indicates an overall
few percent agreement and a particularly good 
match on the low $W^2$ side of the peak, which is important for the
determination of the higher moments.

At large $Q^2$ values, the parameterization from Ref.~\cite{NMC97_f2_fit}
fails to reproduce the data at high $x$. To correct the parameterization
in this
kinematic domain we switched to an older version of the fit reported
in Ref.~\cite{NMC95_f2_fit} for $x>0.75$. However, in order to match the two fits
we had to replace the $x$ variable in the parameterization from
Ref.~\cite{NMC95_f2_fit} with $x^\prime=x-0.9 (x-0.75)^2$.

\section{Fit of the ratio $R\equiv \sigma_L / \sigma_T$}\label{app:rlt_model}
The ratio $R(x,Q^2)$ for the proton is well established in the DIS region and
can be fairly well described by the SLAC fit from Ref.~\cite{r_fit_dis}.
However, until recently, experimental data in the resonance region were missing.
The data from Hall C published in Ref.~\cite{Keppel_R} cover the entire
resonance
region and extend down to very low $Q^2$ values. It was shown by the HERMES
Collaboration in Ref.~\cite{HERMES_R} that in DIS the ratio $R$ does not
depend on the nuclear mass number $A$. We take this assumption to be valid
also in the resonance region. But since smearing effects of Fermi motion
were expected to change both the $F_2$ and $F_L$ structure functions,
we performed a ``smooth'' parameterization of the measured ratio $R$.
This smooth parameterization is based on the fit from Ref.~\cite{r_fit_dis},
modified at low $Q^2$ values and large $x$ by means of a multiplicative
factor:
\begin{eqnarray}
&& \Phi_Q=\Biggl (\frac{Q^2}{Q_0^2}\Biggr )^{C_Q}
\exp{\Bigl[-B_Q C_Q \Bigl(\frac{Q^2}{Q_0^2}-1\Bigr)\Bigr]} \\ \nonumber
&& \Biggl (1-\frac{W_{th}^2-C_W^2}{W^2} \Biggr )^{B_W} ~~,
\end{eqnarray}
\noindent with $Q_0^2=0.8$ (GeV/c)$^2$, $C_Q=0.729$, $B_Q=2.14$, $C_W=0.165$ GeV,
$B_W=0.383$ and where
$W_{th}=M+m_\pi$ is the value of the invariant mass at the pion threshold.
Parameters listed above were obtained from the fit to experimental
data on $R$ taken from Refs.~\cite{Keppel_R,Drees}.

In this way the resonance structures, clearly seen on the proton,
were averaged out to a mean curve. The difference between these two models
gave us an estimate of the systematic uncertainties of the ratio $R$, which
turned out to be very small.

\begin{figure}[!h]
\centering
\includegraphics[bb=2cm 4cm 18cm 24cm, scale=0.4]{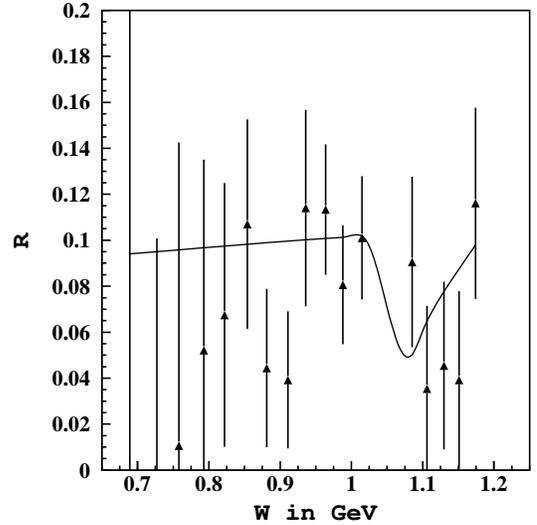}
\caption{Deuteron ratio $R=\sigma_L/\sigma_T$ in the quasi-elastic region
as a function of $W$ at $Q^2=3.25$ (GeV/c)$^2$. The points are from
Ref.~\cite{lung_rlt_qe}
and the curve represents the parameterization described in text.
The minimum at $W=1.07$ GeV is due to the pion electroproduction threshold.}
\label{fig:c_rlt1}
\end{figure}

The ratio $R$ under the quasi-elastic peak is a separate issue. Because of
the nature of the quasi-elastic peak, $R$ is no longer independent of $A$ and
should therefore be treated
within a nuclear model calculation. We used the model from
Ref.~\cite{Simula_QE},
which treats separately the longitudinal and transverse nuclear response
functions
$R_L$ and $R_T$ to obtain the ratio $R$. The conventional ratio $R$ can be
calculated from those quantities as follows:
\begin{equation}\label{eq:c_rlt1}
R=\frac{2Q^2}{Q^2+\nu^2}\frac{R_L}{R_T}
\end{equation}
The deuteron quasi-elastic ratio $R$ obtained from this model
was compared to the data
on the ratio $R$ for the deuteron~\cite{lung_rlt_qe} (see
Fig.~\ref{fig:c_rlt1})
and other nuclei~\cite{r_qe_nucl} (deuteron data on $R$ in the quasi-elastic
region are scarce).
Furthermore, the calculations were compared to the sum of the proton and
neutron form factors, which simply implies:
\begin{equation}\label{eq:c_rlt2}
R=\frac{G_E^2}{\tau G_M^2} ~~,
\end{equation}
\noindent where $\tau=Q^2/4M^2$ and $G_E$, $G_M$ are sums of the known
Sachs form-factors of
the proton and neutron. Since the number of protons and neutrons is different
in different nuclei we did not expect to have the same ratio $R$ for all of
them.
At the same time the $x$-shape of the ratio is very similar from nucleus to
nucleus. In the low $Q^2$ region, where precise data exist, the
calculations reproduce the $x$-shape of the data reasonably well.
The systematic uncertainty was estimated as a difference between the
model calculation
and the result of the naive form-factor sum given by Eq.~\ref{eq:c_rlt2}.

\end{document}